\documentclass[iop]{emulateapj}
\usepackage{natbib}
\usepackage{graphicx}
\usepackage{multirow}

\usepackage[T1]{fontenc}
\usepackage[english]{babel}
\usepackage{array}
\usepackage{textcomp}
\usepackage{url}
\usepackage{amstext}
\usepackage{amssymb}
\usepackage{graphicx}
\usepackage{booktabs}

\newcommand{\lum}{{\rm erg\,s^{-1}}}
\newcommand{\flux}{{\rm erg\,s^{-1}cm^{-2}}}

\newcommand{\xl}{$L_X$}

\newcommand{\xmm}{\textit{XMM-Newton }}

\newcommand{\mic}{$\mu$m}                      %
\newcommand{\wise}{$\it WISE$ }                      %
\newcommand{\nwise}{$\it WISE$}                      %
\newcommand{\buxs}{{\tt BUXS} }
\newcommand{\nbuxs}{{\tt BUXS}}

\newcommand{\cv}{$f_{\rm 2}$ }
\newcommand{\ncv}{$f_{\rm 2}$}

\shorttitle{AGN torus covering factors}
\shortauthors{Mateos et al.}

\begin{document}


\title{X-ray absorption, nuclear infrared emission and dust covering
  factors of AGN: testing Unification Schemes}


\author{S. Mateos\altaffilmark{1},
  F.~J.~Carrera\altaffilmark{1},
  A.~Alonso-Herrero\altaffilmark{1},
  A.~Hern\'an-Caballero\altaffilmark{1},
  X.~Barcons\altaffilmark{1},
  A.~Asensio Ramos\altaffilmark{2,3},
  M.~G.~Watson\altaffilmark{4},
  A.~Blain\altaffilmark{4},
  A.~Caccianiga\altaffilmark{5},
  L.~Ballo\altaffilmark{5},
  V.~Braito\altaffilmark{6},
  C.~Ramos Almeida\altaffilmark{2,3}
}

\altaffiltext{1}{Instituto de F\'isica de Cantabria (CSIC-Universidad de Cantabria), 39005, Santander, Spain; E-mail: mateos@ifca.unican.es}
\altaffiltext{2}{Instituto de Astrof\'isica de Canarias, 38205, La Laguna, Tenerife, Spain}
\altaffiltext{3}{Departamento de Astrof\'isica, Universidad de La Laguna, 38206, La Laguna, Tenerife, Spain}
\altaffiltext{4}{Physics and Astronomy, University of Leicester, University Road, Leicester LE1 7RH, UK}
\altaffiltext{5}{INAF-Osservatorio Astronomico di Brera, via Brera 28, I-20121 Milano, Italy}

\altaffiltext{6}{INAF-Osservatorio Astronomico di Brera, Via Bianchi 46, I-23807 Merate (LC), Italy}


\begin{abstract}
We present the distributions of geometrical covering factors of active
galactic nuclei (AGNs) dusty tori (\ncv) using an X-ray selected
complete sample of 227 AGN drawn from the Bright Ultra-hard \xmm
Survey. The AGN have $z$ from 0.05 to 1.7, 2-10 keV luminosities
between 10${\rm ^{42}}$ and 10${\rm ^{46}}$ $\lum$ and Compton-thin
X-ray absorption. Employing data from UKIDSS, 2MASS and the Wide-field
Infrared Survey Explorer in a previous work we determined the
rest-frame 1-20\,\mic\, continuum emission from the torus which we
model here with the clumpy torus models of Nenkova et al. Optically
classified type 1 and type 2 AGN are intrinsically different, with
type 2 AGN having on average tori with higher \cv than type 1
AGN. Nevertheless, $\sim$20 per cent of type 1 AGN have tori with
large covering factors while $\sim$23-28 per cent of type 2 AGN have
tori with small covering factors. Low \cv are preferred at high AGN
luminosities, as postulated by simple receding torus models, although
for type 2 AGN the effect is certainly small. \cv increases with the
X-ray column density, which implies that dust extinction an X-ray
absorption takes place in material that shares an overall geometry and
most likely belongs to the same structure, the putative torus. Based
on our results, the viewing angle, AGN luminosity and also \cv
determine the optical appearance of an AGN and control the shape of
the rest-frame $\sim$1-20\,\mic\, nuclear continuum emission. Thus,
the torus geometrical covering factor is a key ingredient of
unification schemes.
\end{abstract}


\section{Introduction}
The simplest standard unified models postulate that the diversity of
observed properties of active galactic nuclei (AGNs) can be largely
explained as a viewing angle effect and anisotropic nuclear
obscuration (\citealt{antonucci93}; \citealt{urry95}). A key
ingredient of these orientation-based models is an optically- and
geometrically-thick toroidal structure located on tens of parsec
scales that obscures the AGN nuclear region (accretion disk and X-ray
corona) and the broad line region from certain lines-of-sight. For the
sake of simplicity we will refer to this structure as the
`torus'. Orientation-based unified models of tori with homogeneous
dust distributions propose that AGN are optically classified as type 1
if they are observed at low inclinations with respect to the axis of
the torus. In this case the line of sight does not intercept the
material in the torus and we have an unobscured view of the central
engine. On the other hand, AGN are optically classified as type 2 if
they are observed at high inclinations where the material in the torus
does intercept the line of sight obscuring the central engine (see
\citealt{netzer15} for a recent review).

In recent years it has been realized that the AGN luminosity had to be
incorporated as a key ingredient of unified models to explain the
observed decrease in the relative fraction of type 2 objects at high
AGN luminosities. Such a trend, mainly detected in X-ray surveys, has
been further confirmed by surveys at optical and infrared wavelengths
(e.g. \citealt{hasinger05}; \citealt{simpson05}; \citealt{ceca08};
\citealt{treister08}; \citealt{ebrero09}; \citealt{burlon11};
\citealt{ueda14}; \citealt{buchner15}; \citealt{assef15};
\citealt{lacy15}). To explain the scarcity of luminous type 2 AGN the
'receding torus model' has been often invoked. This model postulates
that the luminosity-dependence of the type 2 AGN fraction is directly
associated with the geometry of the torus in the sense that the
covering factor of the torus (the fraction of the sky as seen by the
source obscured by dust) decreases with increasing AGN luminosity
(\citealt{lawrence91}). Unfortunately, it is yet not fully understood
how the AGN accretion power can influence the physical extent of the
torus as there is substantial quantitative disagreement between
published luminosity trends (e.g. \citealt{lawrence10};
\citealt{sazonov15}).

The nuclear spectral energy distributions (SEDs) and mid-infrared
interferometric observations of nearby AGN are modelled better with
clumpy dusty tori, i.e. the obscuring material is not uniformly
distributed inside the torus. In fact the material appears to be
distributed in discrete, optically thick clumps
(e.g. \citealt{alonso03}; \citealt{tristram07};
\citealt{markowitz14}). X-ray spectral variability studies have also
confirmed that the gas responsible for most of the X-ray absorption
must be clumpy (e.g. \citealt{risaliti09}; \citealt{brenneman13}).

The clumpy nature of the dusty torus has very important implications
for unification models as the classification of an AGN turns out to be
an orientation-dependent probability. This means that even if we
observe from an AGN's equator there is some probability of classifying
it as type 1, while a pole-on AGN could still be classified as type 2
if a single cloud intercepts the line of sight. Thus, while in the
simplest orientation-based models with a smooth torus type 1 and type
2 AGN should have on average tori sharing the same properties, in the
framework of `Clumpy Unification' type 2 AGN should have tori with
geometrical covering factors higher, on average, than type 1 AGN (for
a recent review on this topic see \citealt{elitzur12}). Recent
analyses of the nuclear infrared emission of AGN with radiative
transfer models of clumpy tori indicate that this might be indeed the
case (\citealt{ramos-almeida11}; \citealt{ichikawa15}). Unfortunately
until very recently such studies have been restricted to small samples
of mostly nearby Seyfert galaxies and quasars (\citealt{mor09};
\citealt{nikutta09}; \citealt{ramos-almeida09,ramos-almeida11};
\citealt{alonso11}; \citealt{deo11}; \citealt{lira13}).

Thanks to the advent of the all-sky infrared survey conducted with the
Wide Field Infrared Survey Explorer at 3.4, 4.6, 12 and 22\,\mic\,
(\nwise; \citealt{wright10}) it is now possible to constrain the
properties of the AGN tori, in particular its geometrical covering
factor, in large, representative samples of objects spanning a broad
range of both redshifts and AGN luminosities.

The aim of this study is to verify observationally the validity of
unified schemes in the framework of clumpy torus models. To do so we
have determined, for the first time, the distribution of covering
factors of AGN tori using a large, uniformly selected, complete sample
of AGN. We have investigated whether type 1 and type 2 AGN are indeed
intrinsically different objects, as recently claimed in the
literature, by comparing the distributions of covering factors of
their tori. Finally we have determined the dependence (or lack of) of
the torus covering factor on the line of sight absorption measured in
X-rays and the AGN luminosity.

The 227 AGN used in this study are drawn from the Bright Ultra-hard
\xmm Survey (\nbuxs; \citealt{mateos12,mateos15}; hereafter M15). The
objects have $z$ in the range 0.05-1.7, intrinsic
(absorption-corrected) 2-10 keV X-ray luminosities between 10${\rm
  ^{42}}$ and 10${\rm ^{46}}$ $\lum$ and X-ray absorption in the
Compton-thin regime. There are a number of reasons why we have chosen
the \buxs survey to conduct this study. Firstly, its large sample size
and high spectroscopic identification completeness: \buxs is one of
the largest (255 objects) complete flux-limited samples of bright AGN
selected at energies above 4.5 keV with the \xmm observatory. At the
time of writing optical spectroscopic classifications and accurate
redshifts are available for 98.4 per cent of the objects. Secondly,
the rich set of multiwavelength data available for all sources: good
quality \xmm spectroscopy is available for the full sample, enabling
accurate measurements of both the X-ray absorption and intrinsic X-ray
luminosities for all objects. Furthermore, 227 out of 233 AGN with
X-ray luminosities and redshifts in the chosen intervals have
rest-frame 1-20\,\mic\, nuclear photometric SEDs associated with the
emission from the AGN dusty torus from M15. Clearly, all these
properties make our AGN sample ideally suited to draw robust
statistical constraints on the properties of the dusty torus of AGN.

From all the radiative transfer codes available in the literature to
model the infrared emission associated with clumpy tori
(e.g. \citealt{dullemond05}; \citealt{nenkova02} henceforth referred
to collectively as N08; \citealt{schartmann08}; \citealt{honig10};
\citealt{stalevski12}; \citealt{siebenmorgen15}) we have chosen the
N08 models, referred to as {\sc CLUMPY} models, as they provide a good
representation of the torus SED, and facilitate direct comparison with
previous results in the literature (\citealt{mor09};
\citealt{nikutta09}; \citealt{alonso11}; \citealt{deo11};
\citealt{ramos-almeida09,ramos-almeida11}; \citealt{lira13}).

This paper is structured as follows. Section 2 describes the AGN
sample used in this study. In Sections 3.1 and 3.2 we present the SED
fitting techniques used to first isolate the emission associated with
the torus and then to model it with the N08 models. In Section 3.3 we
discuss our approach to determine the distributions of covering
factors of AGN tori. In Section 4 our results are presented and
discussed while in Section 5 we summarize our main results. Throughout
this paper, errors are 68 per cent confidence for a single parameter,
and we assume $\Omega_{\rm M}$=0.3, $\Omega$${\rm _\Lambda}$=0.7 and
$H_{\rm 0}$=70 km s$^{\rm -1}$ Mpc$^{\rm -1}$.

\section{AGN sample} \label{agn_sample}
The AGN in this study are drawn from the wide-angle Bright Ultra-hard
\xmm Survey ({\tt BUXS}). {\tt BUXS} is a complete flux-limited sample
of 255 X-ray bright AGN (${\tt {\it f}_{4.5-10\,keV} > 6\,x\,10^{-14}
  \flux}$) detected at 4.5 to 10 keV energies with the \xmm European
Photon Imaging Camera (EPIC)-pn (\citealt{struder01}). The objects
were selected at such high energies to reduce as much as possible
biases against highly absorbed AGN. The survey covers 44.43 deg$^2$ of
the northern sky (galactic latitudes $|b|$$>$20 $\deg$) distributed
over 381 \xmm observations having good quality for serendipitous
source detection (\citealt{mateos08,mateos12}). For full details on the survey
design, sample selection and UV/optical spectroscopic identification
and classification of the objects see \citet{mateos12,mateos15}. At
the time of writing the identification completeness is 98.4\,per\,cent
(251 objects). Such a high identification rate guarantees that our
study will not suffer from biases associated with optical
identification incompleteness, that are more severe for highly
obscured type 2 AGN.
 
We have good-quality \xmm spectra for all \buxs sources, which
constrain directly both the line of sight rest-frame absorbing column
densities and X-ray luminosities (see M15). Throughout this paper \xl\,
represents intrinsic, absorption corrected luminosities in the
rest-frame 2-10 keV band.

For the analysis presented here we selected the 233 non-blazar AGN
with $L_X$$>$${\rm 10^{42}}$ and $z$$<$1.7. The luminosity cut was applied
to reduce to a minimum the uncertainties associated with the
determination of the infrared emission of the tori of our objects by
increasing the contrast of the AGN over the underlying emission from
the AGN hosts (12 objects removed; see Sec.~\ref{nir}). The redshift
cut was imposed to assure adequate wavelength sampling of the torus
rest-frame continuum emission (six objects removed). Finally, we
excluded five type 1 AGN and one type 2 AGN without detections with
signal-to-noise-ratio (SNR)$>$2 at all 3.4, 4.6 and 12\,\mic\, in the
final data release of \wise (AllWISE; \citealt{cutri13}). As the
number of objects not detected with \wise above our selection
threshold is rather small and, in addition, these objects span a broad
range of \xl\, and $z$, removing them from the sample should not bias
our results. All the above selection criteria left us with a sample of
227 AGN with $L_X$ from ${\rm 10^{42}}$ to ${\rm 10^{46}\lum}$ and
0.05$<$z$<$1.7.
 
We classified our AGN as type 1 if permitted and semi-forbidden broad
emission lines (line velocity widths $\gtrsim$1500\,km\,s$^{\rm -1}$)
were detected in their rest-frame UV/optical spectra (132 objects) and
as type 2 if they showed narrow emission lines only (line velocity
widths $<$1500\,km\,s$^{\rm -1}$; 75 objects) or had a galaxy-like
spectrum with no emission lines (3 objects). Due to the controversy
regarding the nature of intermediate Seyfert types 1.8 and 1.9 as type
1 or type 2 AGN we kept such objects as a separate class (17
objects)\footnote{Objects of intermediate Seyfert type 1.9 can be
  identified up to $z$$\sim$0.2-0.4, depending on whether the
  H$\alpha$ emission line is outside the observable wavelength range
  of our spectroscopic data or not.}.

\begin{figure}[!t]
  \centering
  \begin{tabular}{cc}
    \hspace{-0.7cm}\includegraphics[angle=90,width=0.50\textwidth]{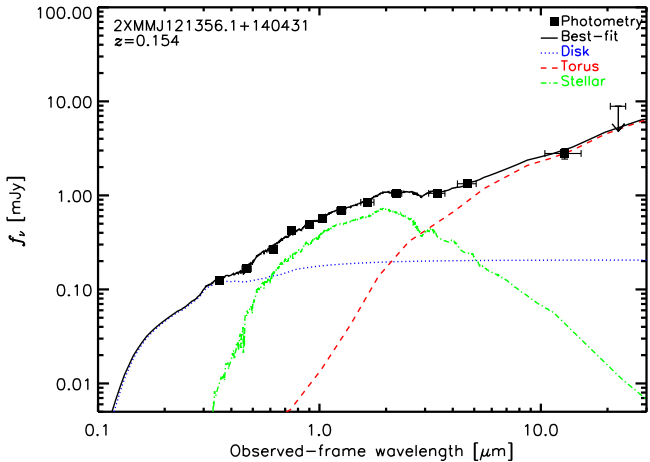}\\
  \end{tabular}
  \caption{Example of the SED decomposition analysis used in M15 to
    isolate the AGN dusty torus emission (see Sec.~\ref{nir} for
    details). The example corresponds to an AGN optically classified
    as type 1. Filled squares are the catalogued photometry. The
    dotted, dashed and dotted-dashed lines correspond to the accretion
    disk, torus (Seyfert 1 template from \citet{silva04}) and host
    galaxy emission, respectively. The solid line is the best-fit
    model. We note that in the SED decomposition analysis we treated
    all 22\,\mic\, detections (168 in total) as upper limits (see M15
    for details).}
    \label{fig1}
\end{figure}

\section{Methodology}
\subsection{Isolating the torus emission} \label{nir}
In M15 we determined infrared SEDs associated with the
emission from the dusty torus for the AGN in \nbuxs. To do so we
conducted a thorough analysis of the rest-frame UV-to-infrared
photometric SEDs to correct the catalogued infrared fluxes for any
contamination associated with both the host galaxies and the direct
emission from the AGN accretion disk. Our SEDs are based on data from
the Sloan Digital Sky Survey (SDSS; \citealt{abazajian09}), the Two
Micron All Sky Survey (2MASS; \citealt{jarrett00}; \citealt{cutri13}),
the UKIRT Infrared Deep Sky Survey (UKIDSS; \citealt{lawrence07}) and
WISE (\citealt{wright10}). To decompose the observed fluxes into AGN
and galaxy emission we used the SED fitting tool SEd Analysis using
BAyesian Statistics
(SEABAS\footnote{http://astro.dur.ac.uk/$\sim$erovilos/SEABASs/},
\citealt{rovilos14}).

Very briefly, to model the emission from the accretion disk we used
the type 1 quasar SED from \citet{richards06} at rest-frame
wavelengths $\lambda$$<$0.7\mic\, and a power-law ${\rm \lambda}
f_{\rm \lambda} \propto {\rm \lambda}^{\rm -1}$ at longer
wavelengths. To redden the accretion disk we used the \citet{gordon98}
Small Magellanic Cloud extinction law at $\lambda$<0.33\,\mic\, and
the \citet{cardelli89} Galactic extinction law at
$\lambda$$>$0.33\,\mic. In both cases we assumed $R_V$=3.1. To
characterize the continuum emission from the AGN dusty torus we used
the Seyfert 1 and the two Seyfert 2 templates corresponding to
rest-frame X-ray absorbing column densities ${\rm N_H}$$<$${\rm
  10^{24}\,cm^{-2}}$ from \citet{silva04}. Finally, to reproduce the
emission from the stellar population of the AGN hosts at rest-frame
optical-near-infrared wavelengths we used a library of 75 stellar
templates from \citet{bruzual03}. The templates have solar metallicity
and a Chabrier initial mass function (\citealt{chabrier03}) and were
generated using 10 exponentially decaying star formation histories
with characteristic times $\tau$=0.1-30\,Gyr and a model with constant
star formation, and a set of ages in the range 0.1-13\,Gyr. To redden
the stellar templates we used the \citet{calzetti00} dust extinction
law. An example of the SED decomposition analysis is illustrated in
Fig.~\ref{fig1}. It is important to highlight that we have adopted the
same templates and SED-decomposition procedure to isolate the AGN
torus emission of all sample objects.

In M15 we demonstrated that stellar contamination of catalogued fluxes
in the infrared regime at rest-frame wavelengths shorter than
$\sim$6\,\mic\, is significant, especially for type 2 objects. Only
for type 1 objects with \xl$>$$10^{44}$\,$\lum$ the AGN outshines the
host galaxy in the infrared band. The tight correlation found between
rest-frame 6\,\mic\, luminosities, corrected for contamination from
the accretion disk and AGN hosts, and 2-10 keV intrinsic (absorption
corrected) luminosities, supports the hypothesis that the infrared
SEDs determined from our decomposition analysis are associated with
dust heated by the intense radiation field of the AGN. This dust is
most likely located in the putative torus on parsec scales. Hereafter,
nuclear infrared SEDs will refer to the emission from the torus.

We focus our analysis here on rest-frame wavelengths longer than
1\mic\, since this is the spectral region where the AGN torus emits
the bulk of its radiation. Although in M15 we demonstrate that, at the
luminosities of our AGN, contamination due to star formation at these
wavelengths should be negligible, to minimize such effect we have
treated the \wise 22\,\mic\, fluxes of all 12 objects with
\xl$<$5$\times$${\rm 10^{42}}\lum$ (five type 1 AGN, six type 2 AGN
and one Sy1.9), whether detected or not at these wavelengths, as upper
limits. Nevertheless, we have checked that this assumption does not
affect our main results.

\begin{figure}[!h]
  \centering
  \begin{tabular}{cc}
    \hspace{-0.7cm}\includegraphics[angle=90,width=0.50\textwidth]{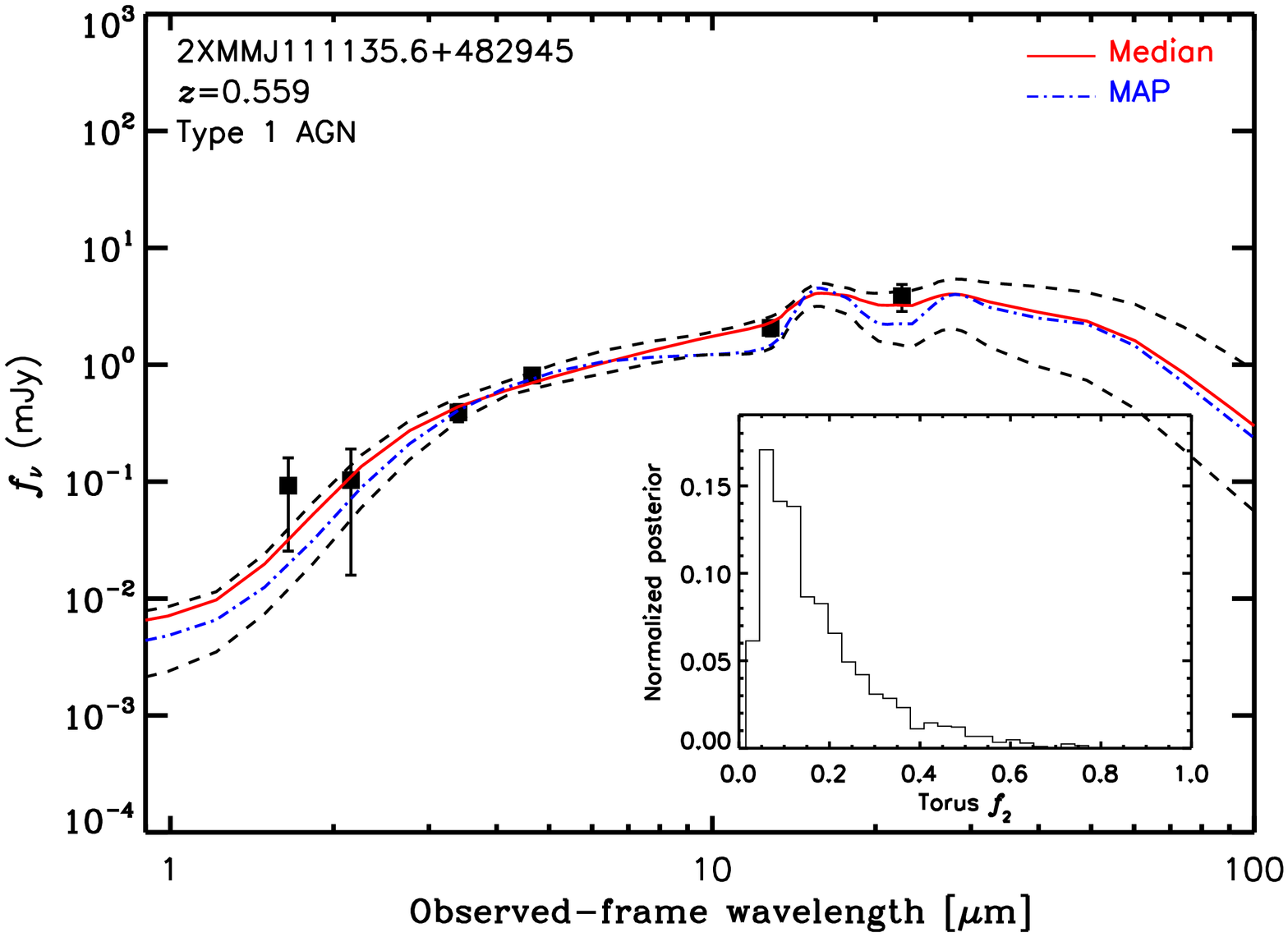}\\
    \hspace{-0.7cm}\includegraphics[angle=90,width=0.50\textwidth]{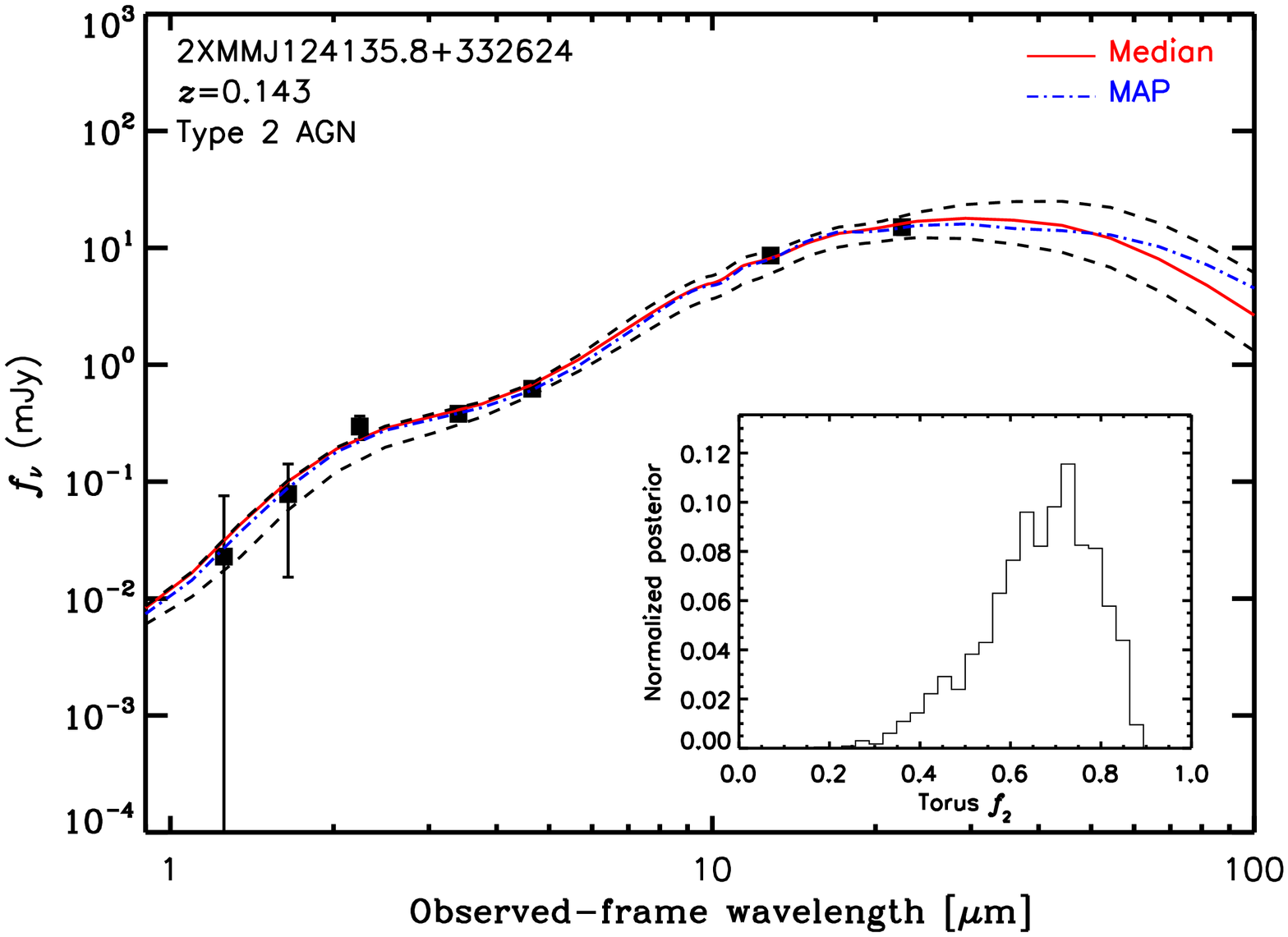}\\
        \hspace{-0.7cm}\includegraphics[angle=90,width=0.50\textwidth]{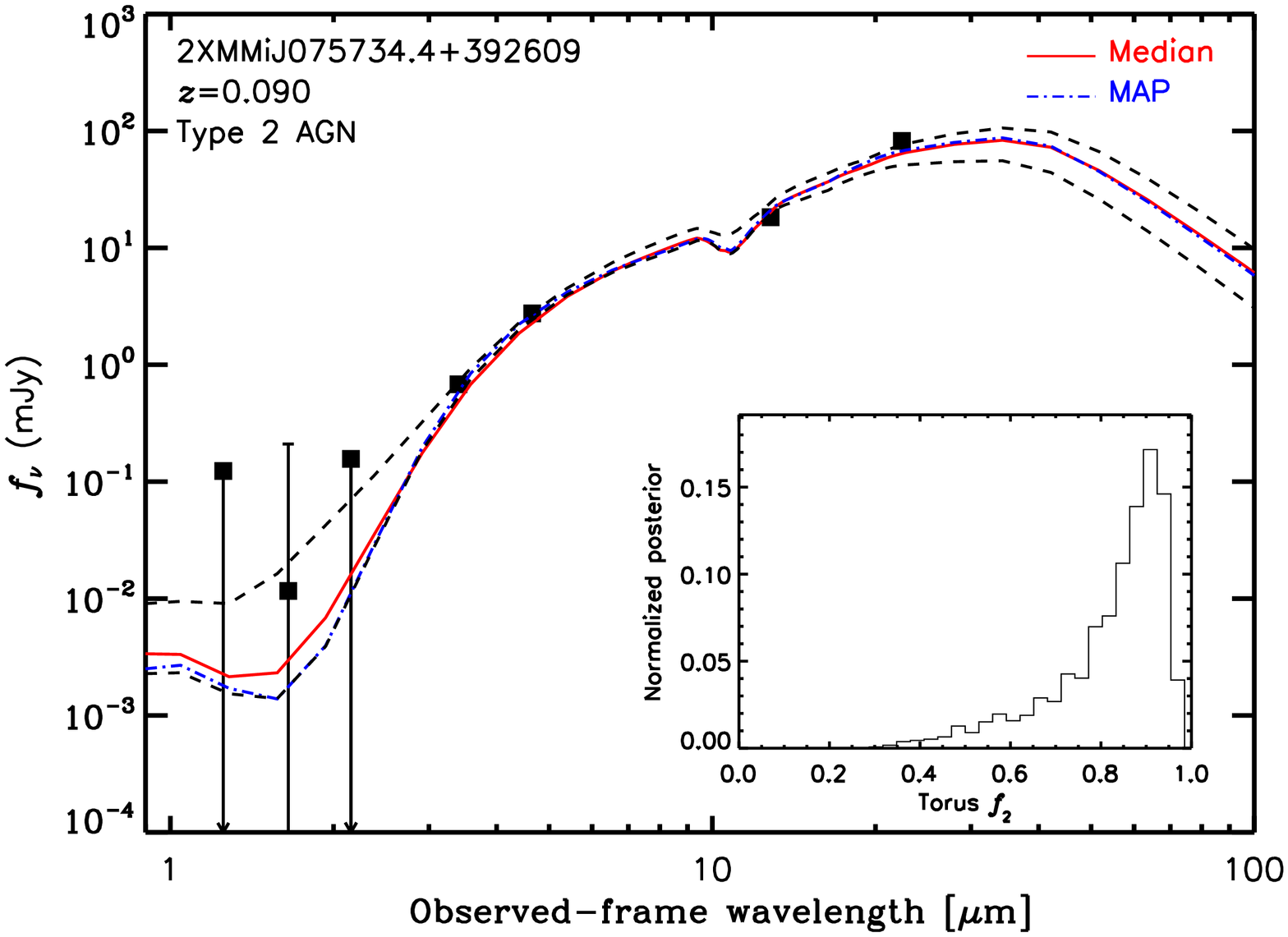}\\

  \end{tabular}
  \caption{Examples of the SEDs of AGN dusty tori used in our study
    (filled symbols and error bars). Vertical arrows indicate upper
    limits. The SEDs have been corrected for the emission associated
    with both the accretion disk and the AGN host galaxies. The solid
    and dot-dashed lines correspond to the torus models described with
    the maximum (mode-MAP) and median values of the posterior
    probability distributions of the parameters returned by the
    Bayes{\sc CLUMPY} fits, respectively. The dashed lines indicate
    the range of models enclosing a 68 per cent probability. The inset
    plots show the normalized posterior distributions of $f_{\rm 2}$
    from the fits.}
    \label{fig2}
\end{figure}

\subsection{SED fitting with {\sc CLUMPY} models}
To describe the nuclear infrared emission of our objects we have used
the radiative transfer models by N08. In these models the material
obscuring the AGN nuclear region is treated as a medium with a
toroidal shape where the dust and gas is distributed in high-density
clumps inside it. The angular distribution of clouds has no sharp
cutoff boundary and is described as a Gaussian of width $\sigma$,

\begin{eqnarray}
 {N_{\rm T}(\beta)=N_{\rm 0} e^{(-\beta^2 / \sigma^2)}}
\label{eq1}
\end{eqnarray}
where $N_{\rm T}$ is the line of sight number of clouds,
$\beta$=$\pi$/2-$i$ is the inclination angle of the torus equatorial
plane with respect to the line of sight and $i$ is the viewing angle
from the torus axis. $N_{\rm 0}$ represents the mean number of clouds
along radial equatorial rays. In the N08 models the radial
distribution of clouds is parameterized as a power-law, $N(r) \propto
r^{-q}$, where $N$ is the number of clouds and $q$ the power-law
index. The torus radial thickness ($Y$) is defined as the ratio of the
outer ($R_{\rm o}$) to inner ($R_{\rm d}$) radius of the distribution
of clouds, where $R_{\rm d}$ is set by the AGN luminosity and the dust
sublimation temperature ($\sim$1500 K in the model;
\citealt{barvainis87}). All clouds are assumed to be optically-thick
and with the same optical depth, defined in the $V$ band at
5500{\AA}. The model assumes a standard cold oxygen-rich interstellar
medium (ISM) dust extinction law (\citealt{ossenkopf92}). In addition
to the parameters defining the geometry of the torus and the
properties of the clouds, the scaling factor required to match the
fluxes from the best-fit model to the observed values in the SEDs can
be used as proxy for the AGN bolometric luminosity (see
\citealt{nenkova08b}; \citealt{alonso11}). We refer the reader to N08
for further details on {\sc CLUMPY} models.

\begin{table}
  \center
 \caption{Parameters of the N08 {\sc clumpy} torus models and range of values used in this work.}
 \label{tab:parameters}
 \begin{tabular}{@{}lcccccccc}
   \hline
   \hline
  Parameter & Range \\
  \hline
  Torus angular width ($\sigma$) & [15$^{\circ}$-70$^{\circ}$] \\
  Torus radial thickness ($Y$) & [5-30] \\
  Mean number of clouds along equatorial rays ($N_{\rm 0}$) & [1-15] \\
  Index of the radial distribution of clouds ($q$) & [0-3] \\
  Single cloud optical depth ($\tau_{\rm V}$) & [5-150] \\
  Viewing angle ($i$) & [0$^{\circ}$-90$^{\circ}$] \\
  \hline
 \end{tabular}
 \smallskip
 \end{table}
\smallskip

The online database of {\sc CLUMPY} consists of more than $10^6$
models based on a narrow grid for each torus
parameter\footnote{http://www.pa.uky.edu/clumpy/}. With such a fine
grid we will not be able to distinguish between torus models based on
different combinations of parameters but with differences in the
continuum shape smaller than our SED photometric uncertainties. In
that sense we can say that in our analysis there is a strong
degeneracy in the parameters of {\sc CLUMPY} models. To best deal with
this issue we have conducted the SED fits using the code Bayes{\sc
  CLUMPY} from \citet{asensio09}. This code has been especially
developed to analyze the emission of AGN tori with {\sc CLUMPY} models
using a Bayesian inference approach. Bayes{\sc CLUMPY} uses a
Metropolis-Hastings Markov Chain Monte Carlo (MCMC) sampling technique
to determine posterior distributions for each parameter. To ensure the
continuity of parameters, Bayes{\sc CLUMPY} also interpolates in the
original database of models. In our analysis we have used truncated
uniform prior distributions for all parameters in the ranges listed in
Table~\ref{tab:parameters}.

\subsection{Covering factors of AGN tori}
In torus models with a clumpy distribution of dust the UV/optical
appearance of an AGN depends on the viewing angle and the probability
of intercepting a dusty cloud along our line of sight. Thus, the
geometrical covering factor of the torus should play a fundamental
role in the optical classification of AGN: the smaller the covering
factor, the higher the probability of a direct view of the AGN nuclear
region. Assuming that all individual clouds are optically thick, as in
the N08 models, the probability that light from the AGN at an angle
$\beta$ will escape unaffected from the torus has the form
\begin{eqnarray}
 {P_{\rm esc}(\beta)=e^{-N_{\rm 0} e^{(-\beta^2/\sigma^2)}}}
\label{eq1}
\end{eqnarray}
The geometrical covering factor of the torus representing the fraction
of the sky obscuring the AGN nuclear region, $f_2$, is defined as

\begin{eqnarray}
 {f_{\rm 2}=1-\int_0^{\pi/2} \! P_{\rm esc}(\beta)\,{\rm cos}(\beta)\it{d}\beta}
\label{eq2}
\end{eqnarray}
where $P_{\rm esc}$ is integrated over all angles
(\citealt{mor09}). As $f_{\rm 2}$ is independent of the inclination
angle, it represents the true intrinsic fraction of optically obscured
type 2 objects in the entire AGN population.

Fig.~\ref{fig2} shows three examples of the typical nuclear infrared
SEDs used in our study and the SED-fitting results obtained with
Bayes{\sc CLUMPY}. The insets show the normalized posterior
distributions of \cv derived from the fits. To obtain the distribution
of \cv for a sample of objects fully taking into account the
uncertainties from the fits we first concatenated together the
individual arrays of values of \cv returned from the MCMC analysis for
each object and then we computed the probability distribution of the
combined array of values of \ncv\footnote{While we believe that the
  methodology applied is reliable, a more coherent way to infer the
  global distribution of the covering factor could have been to use a
  hierarchical Bayesian model. Nevertheless, such approach would have
  enjoyed the typical shrinkage of hierarchical modelling.}. To
compare different distributions of \cv we have used the two-sample
Kolmogorov-Smirnov (KS) test and Monte Carlo simulations to determine
the probability of rejecting the null hypothesis that the two samples
are drawn from the same parent population. Each time we run our
simulations we used bootstrap to randomly select 10$^5$ times the
sample objects used to determine the distributions of \ncv.

For illustration purposes throughout this paper we have used a bin
size of 0.03 to represent the distributions of \cv but we stress that
all computations are based on the arrays of values of \ncv. All
distributions are normalized to have an area of one under the curve.

\begin{figure}
  \centering
  \begin{tabular}{cc}
    \hspace{-0.7cm}\includegraphics[angle=90,width=0.5\textwidth]{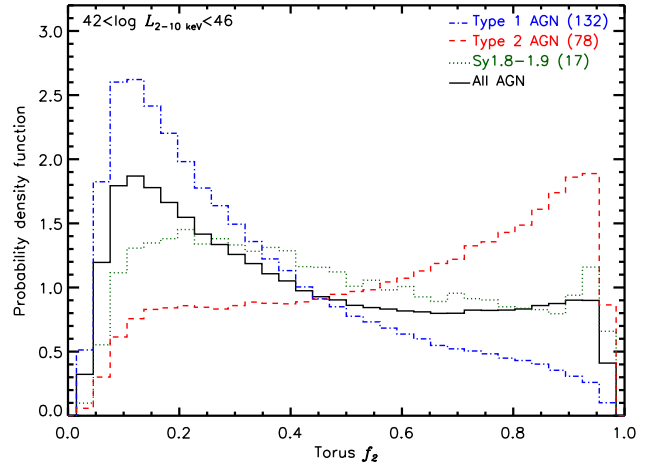}
  \end{tabular}
  \caption{Distributions of the covering factors of AGN tori
    calculated for the full sample of objects.}
  \label{fig3}
\end{figure}

\section{Results}
As indicated before, we have used the N08 models to reproduce the
nuclear infrared emission associated with the dusty tori of AGN and to
determine their dust covering factors. Therefore, the results inferred
from our SED fits should be considered in the framework of these
models.

\subsection{\cv versus optical class}
\label{f2_class}
Fig.~\ref{fig3} shows the distribution of \cv for our full sample of
AGN. It is evident that type 1 and type 2 AGN have significantly
different distributions of \ncv, in the sense that type 2 AGN overall
have tori with higher covering factors than type 1 AGN. Based on the
KS test and our simulation analysis we can reject the null hypothesis
that the two samples are drawn from the same parent population with a
confidence level higher than 99.99 per cent. Nevertheless, we find
that there is a large overlap between the distributions of \cv for
type 1 and type 2 AGN, in good agreement with previous studies based
on high-spatial resolution nuclear infrared photometric data and/or
mid-infrared spectroscopic data for small samples of local Seyferts
and PG quasars (e.g. \citealt{mor09}; \citealt{alonso11};
\citealt{ramos-almeida11, ramos-almeida09};
\citealt{ichikawa15}).

The distributions of \cv that we find for the two AGN populations are
significantly broader than claimed in the above studies. As we will
see in the following sections such apparent discrepancies are not
associated with higher uncertainties in our analysis compared to
previous studies but to type 1 (type 2) AGN having rather large
(small) torus covering factors. Since such objects are rare, sampling
them requires studies of large and complete samples of AGN such as
ours. For example, as we will see in Sec.~\ref{ir_cont_f2_dep}, if we
use \ncv=0.5 as a threshold to separate AGN tori with low and high
covering factors, we find that 26 out of 132 type 1 AGN have tori with
high covering factors while 22 out of 78 type 2 AGN have tori with low
covering factors.

\begin{figure*}[!t]
  \centering
  \begin{tabular}{cc}
    \hspace{-0.7cm}\includegraphics[angle=90,width=0.50\textwidth]{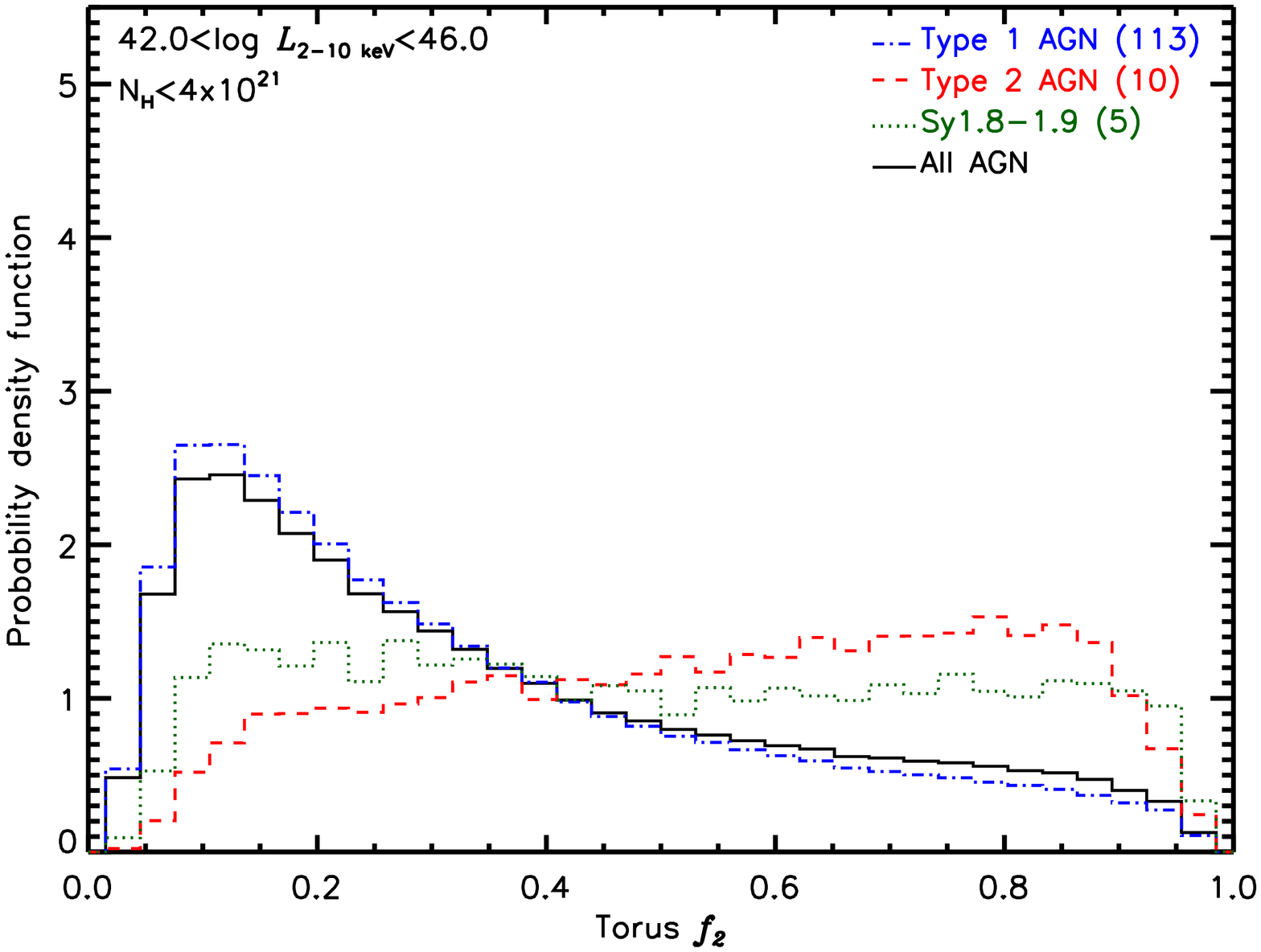}
    \hspace{-0.7cm}\includegraphics[angle=90,width=0.50\textwidth]{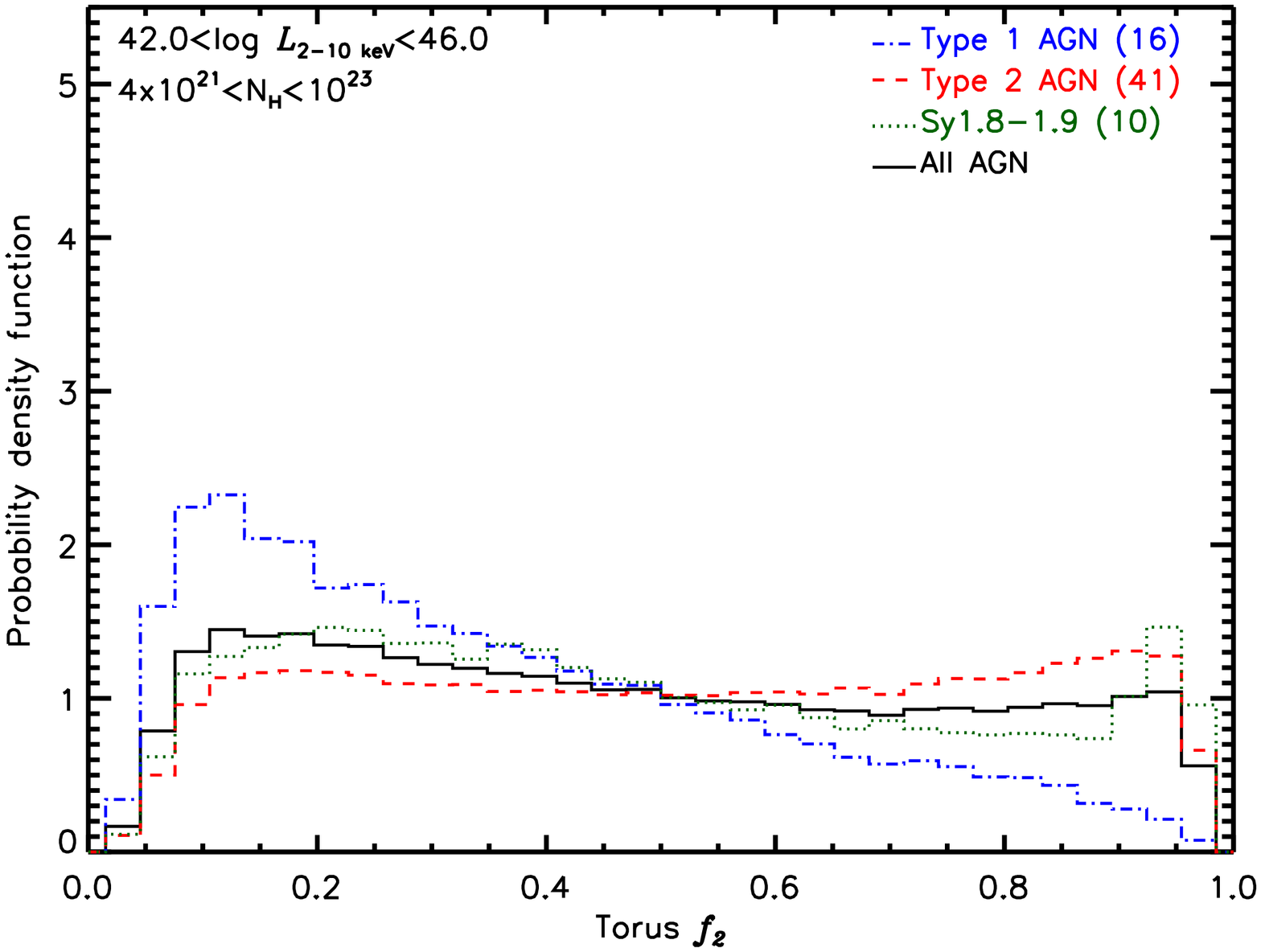}\\
    \hspace{-0.7cm}\includegraphics[angle=90,width=0.50\textwidth]{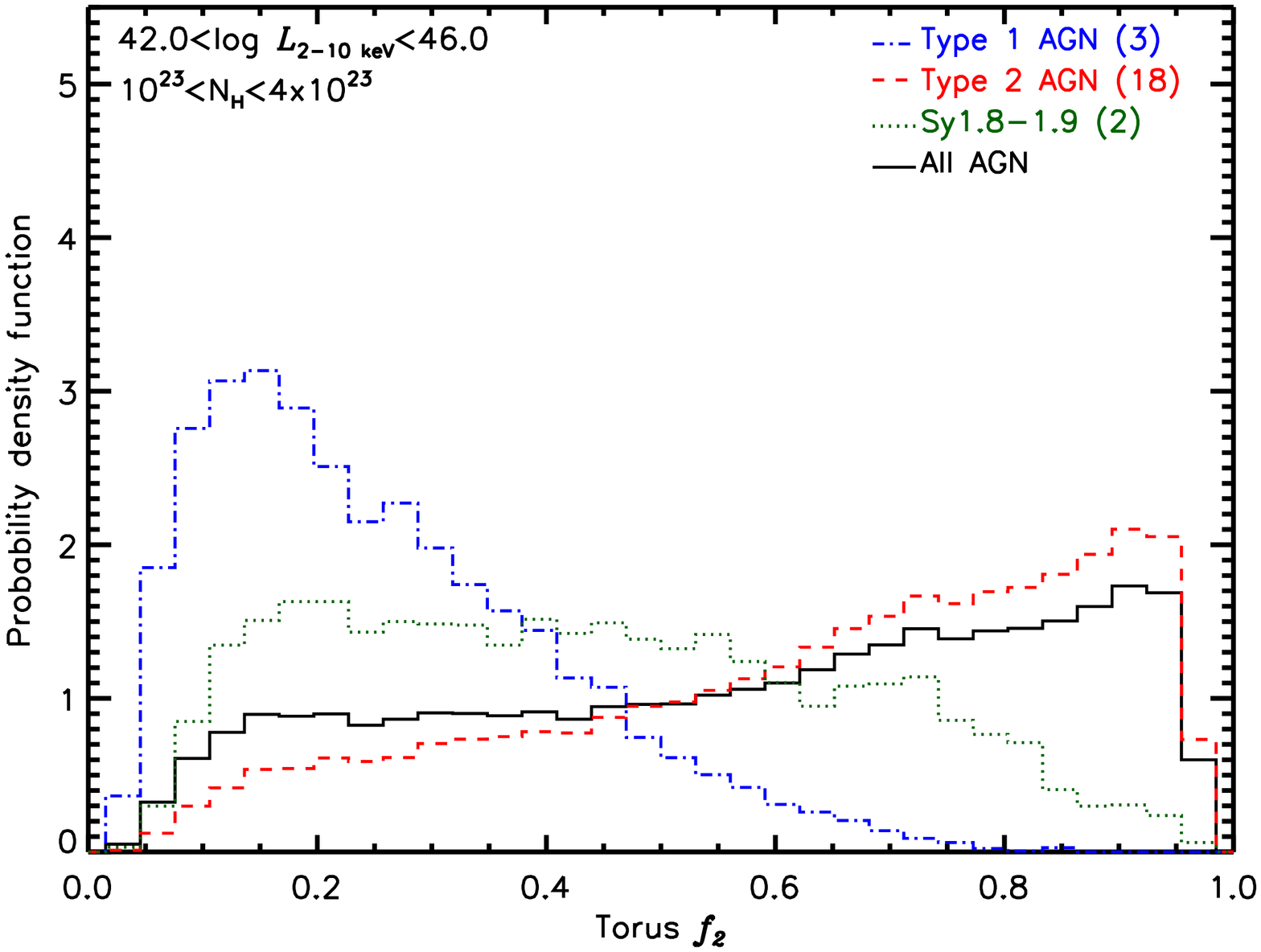}
    \hspace{-0.7cm}\includegraphics[angle=90,width=0.50\textwidth]{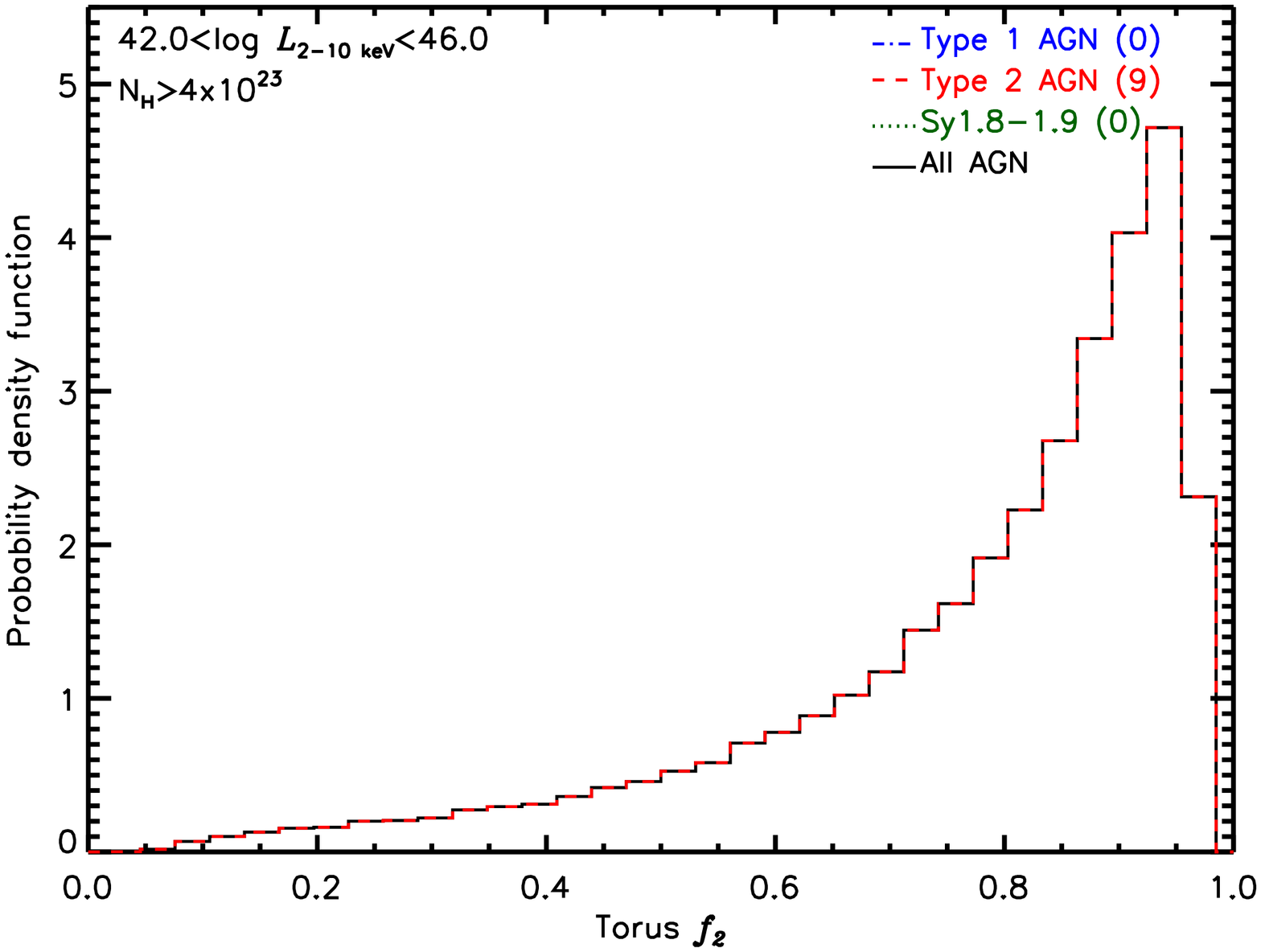}\\
  \end{tabular}
  \caption{Dependence of the distribution of dust covering factors of AGN tori on the line of sight X-ray absorption.}
  \label{fig4}
\end{figure*}

As indicated before, we stress that we have used the same
SED-decomposition procedure to isolate the AGN torus emission of all
sample objects. Our SED fits with Bayes{\sc CLUMPY} take into account
both the quality of the catalogued infrared photometric data, which is
similar among our type 1 and type 2 AGN, and the uncertainties from
the SED decomposition analysis. We have also checked that, the
distributions of \cv for type 1 AGN have widths, based on the 16th and
84th percentiles, indistinguishable from those of type 2 AGN. All this
guarantees that any differences in the torus properties among AGN with
different optical spectroscopic classifications reported here are
genuine and not an artifact of either the SED decomposition analysis
or the quality of the infrared data.

Interestingly, we find that AGN classified as Sy1.8-1.9 have a rather
flat distribution of \cv that is significantly different from those of
type 1 and type 2 AGN. Thus from the infrared point of view, we find
that it is highly unlikely that all Sy1.8-1.9 are simply ordinary type
1 AGN caught in a low flux state during the UV/optical spectroscopic
observations, as significant differences exist between the properties
of the tori of these two AGN classes. A KS test returns probabilities
of 99.78 per cent and 97.98 per cent for rejecting the null hypothesis
that the distributions of \cv for Sy1.8-1.9 and those obtained for
type 1 and type 2 AGN are identical. Therefore, alternative processes
must play a role. For example, our finding that 12 out of the 17
intermediate-type objects are absorbed in X-rays (see Sec.~\ref{f2_nh})
favours a scenario where most of our Sy1.8-1.9 objects have broad-line
regions reddened by optically-thin dust located either in the torus or
on physical scales of the narrow-line region or the host galaxies
(e.g. \citealt{alonso11}).

We note that, if some high-$z$ Sy1.8-1.9 objects are still
present in our sample of type 2 AGN, the effect would be to reduce the
differences between the \cv distributions of type 1 and type 2
AGN. Clearly, this would not change our results since we have already
found that we can reject the hypothesis that the distributions of \cv
for type 1 and type 2 AGN are drawn from the same parent population
with a confidence level higher than 99.99 per cent.

Finally, we have used the results from our SED decomposition analysis
to compare the distributions of \cv for all objects with detected
UV/optical broad emission lines (132 type 1 AGN and 17 Sy1.8-1.9s)
with low and high extinction towards their accretion disk. To separate
the objects we have used an extinction of E(B-V)=0.32, or $A_{\rm
  V}$$\sim$1\,mag assuming a Galactic standard conversion. Such a
value has often been used in the literature to identify moderately
reddened type 1 AGN (e.g. \citealt{urrutia12}; \citealt{lacy13} and
references therein). Based on the chosen extinction threshold, 13 out
of 132 type 1 AGN and 10 out of 17 Sy1.8-1.9s are classified as
moderately reddened objects (E(B-V) in the range 0.32-0.65; see M15
for details). We find that, as expected, higher \cv are preferred in
reddened broad-line AGN. According to the KS test we can reject the
null hypothesis that the \cv distributions of the two samples
(reddened and unreddened broad-line AGN) are drawn from the same
parent population with a 99.3 per cent confidence level.

Based on the results presented in this section we can conclude that,
type 1, type 2 and probably also intermediate-type AGN, are on average
intrinsically different, as has been reported previously in the
literature (e.g. \citealt{ramos-almeida11}).

\subsection{\cv versus X-ray absorption}
\label{f2_nh}
The discovery that the UV/optical spectroscopic classifications of AGN
correlate well with the absorption properties measured in X-rays has
lend strong observational evidence favouring standard
orientation-based unified models. Nevertheless, it is well known that
AGN exhibit a large range of dust-to-gas ratios and that, for a
non-negligible fraction of objects, gas absorption in X-rays and dust
extinction in the UV-to-infrared spectral band are not always
detected together (e.g. \citealt{mainieri02}; \citealt{mateos05a};
\citealt{tozzi06}; \citealt{garcet07}; \citealt{winter09};
\citealt{mateos10}; \citealt{corral11}; \citealt{scott11};
\citealt{page11}; \citealt{omaira14}; \citealt{merloni14}).

To investigate whether a physical or geometrical connection exists
between the material responsible for the X-ray absorption and
UV-to-infrared obscuration, we have computed the distributions of \cv
for AGN with different levels of X-ray absorption. To have a good
representation of both type 1 and type 2 AGN across the full range of
measured X-ray column densities, we have divided the sample in four
different bins: ${\rm N_H}$$<$${\rm 4\times10^{21}\,cm^{-2}}$
(henceforth X-ray unabsorbed), ${\rm 4\times10^{21}}$$<$${\rm
  N_H}$$<$${\rm 10^{23}\,cm^{-2}}$,\,${\rm 10^{23}}$$<$${\rm
  N_H}$$<$${\rm 4\times10^{23}\,cm^{-2}}$ and ${\rm
  4\times10^{23}}$$<$${\rm N_H}$$<$${\rm
  1.4\times10^{24}\,cm^{-2}}$. Fig.~\ref{fig4} summarizes the results
of this analysis. Although in Fig.~\ref{fig4} we show the
distributions of \cv for objects classified as Sy1.8-1.9 for
completeness, we do not use them in the analysis presented in this
section as we are clearly limited by small number
statistics. Nevertheless, our results suggest that the distribution of
\cv is rather flat for both X-ray unabsorbed and absorbed Sy1.8-1.9s.

We note that none of our X-ray selected sources has a best-fit X-ray
column density in the Compton-thick regime. Nevertheless, considering
the uncertainties in ${\rm N_H}$, we cannot rule out unambiguously
Compton-thick absorption in five type 2 AGN (all five sources belong
to the ${\rm 4\times10^{23}}$$<$${\rm N_H}$$<$${\rm
  1.4\times10^{24}\,cm^{-2}}$ bin). As an independent test, we have
determined the $L_X^{\rm obs}$/$L_{6\,\mu {\rm m}}$ luminosity ratio
for these objects where, $L_X^{\rm obs}$ are observed (i.e. not
corrected for intrinsic absorption) rest-frame 2-10 keV luminosities
and $L_{6\,\mu {\rm m}}$ are the monochromatic luminosities of the
torus emission at rest-frame 6\,\mic. The later have been shown to be
a good proxy for the AGN intrinsic power (\citealt{lutz04};
\citealt{ramos-almeida07}; \citealt{fiore09};
\citealt{georgantopoulos11}; \citealt{mateos15};
\citealt{stern15}). Based on the relationship between $L_X$ and
$L_{6\,\mu {\rm m}}$ from M15 we find that, in all five cases, the
$L_X^{\rm obs}$/$L_{6\,\mu {\rm m}}$ ratio is consistent with
Compton-thin absorption. Finally, we have used a column density of
${\rm 4\times10^{21}\,cm^{-2}}$ to separate unabsorbed and absorbed
AGN. Assuming a Galactic standard dust-to-gas ratio, an ${\rm N_H}$ of
${\rm 4\times10^{21}\,cm^{-2}}$ corresponds to A$_{\rm V}$$\sim$2 mag,
or $E(B-V)$$\sim$0.65, the extinction level that separates optical
type 1 from type 2 AGN (\citealt{caccianiga08}; \citealt{merloni14}).

\begin{figure*}[!t]
  \centering
  \begin{tabular}{cc}
    \hspace{-0.7cm}\includegraphics[angle=90,width=0.50\textwidth]{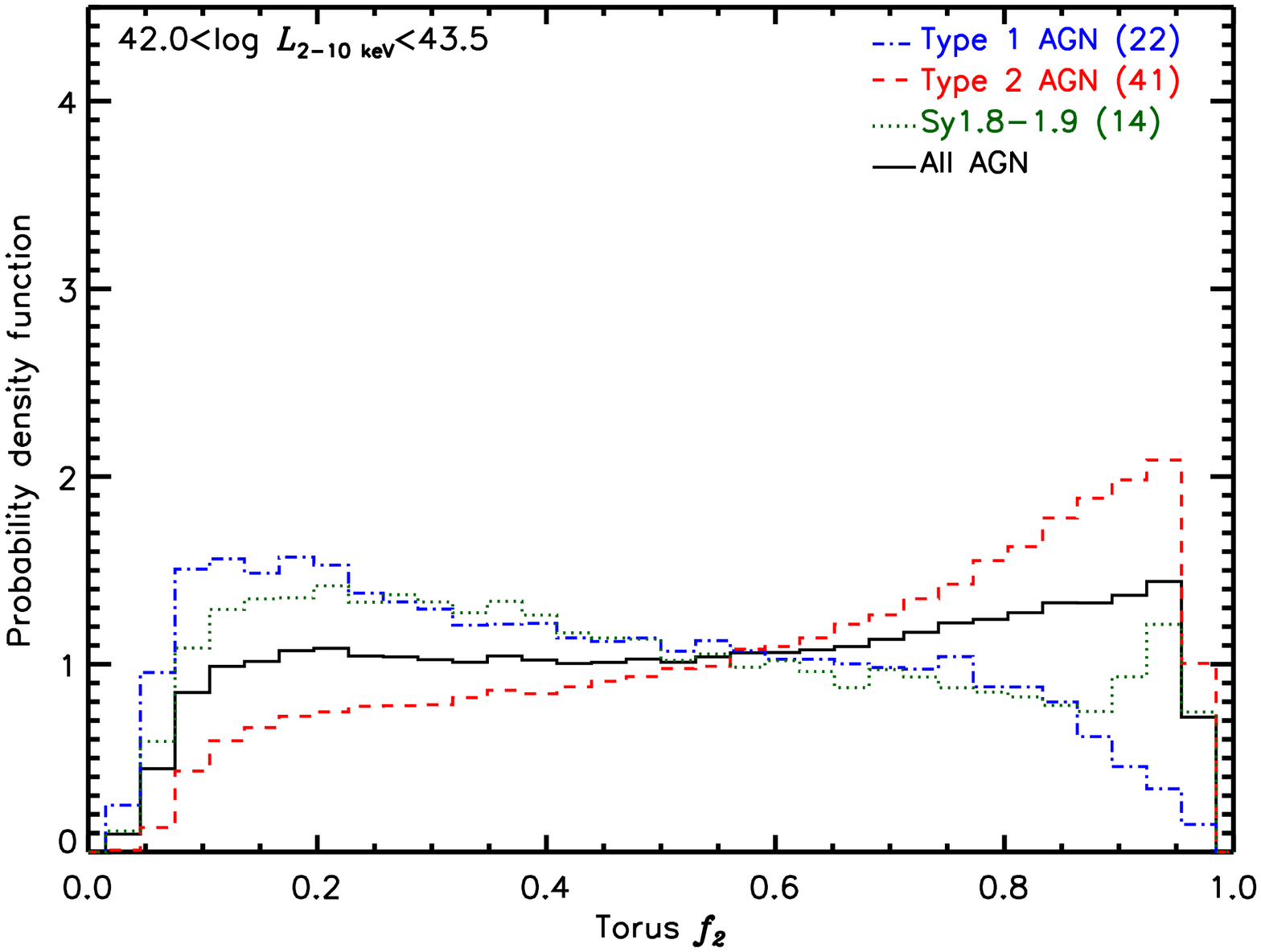}
    \hspace{-0.7cm}\includegraphics[angle=90,width=0.50\textwidth]{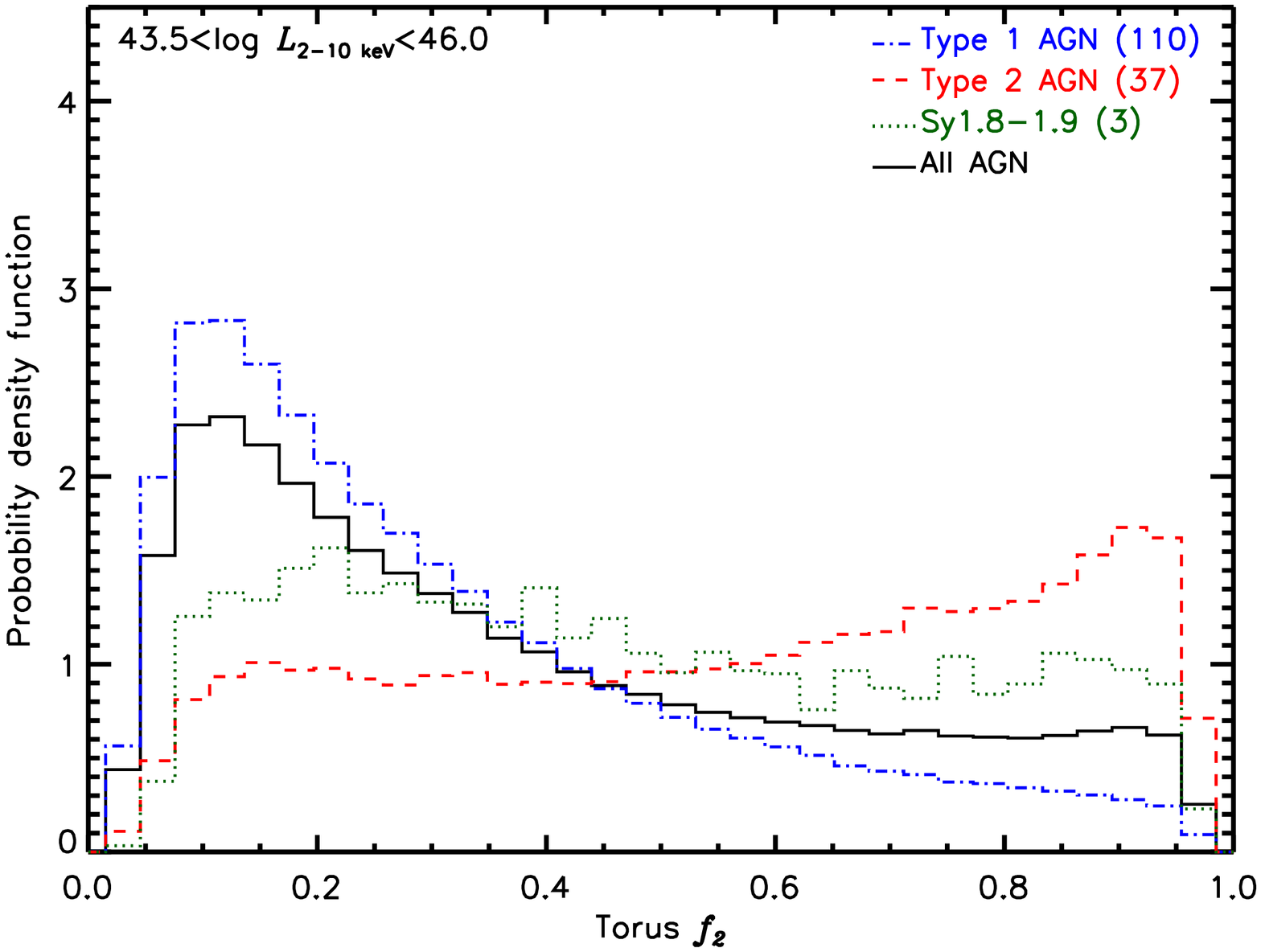}\\
  \end{tabular}
  \caption{Dependence of the distribution of dust covering factors of
    AGN tori on the X-ray luminosity.}
    \label{fig5}
\end{figure*}

Interestingly we find that, type 1 and type 2 AGN with similar levels
of X-ray absorption have significantly different distributions of
torus geometrical covering factors. This result also holds for X-ray
unabsorbed objects (Fig.~\ref{fig4} top left). We can reject the null
hypothesis that the distributions of \cv for X-ray unabsorbed type 1
and type 2 AGN are drawn from the same parent population with a
confidence higher that 99.99 per cent. Clearly, intrinsic differences
exist among the torus properties of these two groups of objects. Thus,
although the host galaxies could totally outshine the AGN emission in
the optical band in some objects (e.g. \citealt{moran02};
\citealt{severgnini03}; \citealt{page06}), this cannot be the sole
factor in determining the optical appearance of X-ray unabsorbed type
2 objects (\citealt{panessa02}; \citealt{bianchi08};
\citealt{gallo13}). Indeed only 4 out of 10 X-ray unabsorbed type 2
AGN in our sample have $L_X$$<$10$^{\rm 43}\,\lum$, where host galaxy
dilution can be an important effect (\citealt{caccianiga07}).

The distributions of \cv for X-ray unabsorbed type 1 and type 2 AGN
(Fig.~\ref{fig4} top left) are largely indistinguishable from those of
absorbed AGN with ${\rm N_{H}}$ in the range ${\rm 4 \times
  10^{21}\,cm^{-2}}$$<$${\rm N_H}$$<$${\rm 10^{23}\,cm^{-2}}$
(Fig.~\ref{fig4} top right). This suggests that up to column densities
of $\sim$10${\rm ^{23}\,cm^{-2}}$ there is no significant correlation
between \cv and ${\rm N_{H}}$. Nevertheless, at ${\rm N_H}$$<$${\rm
  10^{23}\,cm^{-2}}$ gas and dust in the AGN hosts might be
contaminating some of our measurements, especially for type 2 AGN
(e.g. \citealt{alonso03}; \citealt{guainazzi05};
\citealt{goulding12}). Therefore we focus our attention on objects
with column densities ${\rm N_H}$$>$${\rm 10^{23}\,cm^{-2}}$ as such
high column densities should be associated with the
torus\footnote{Typical optical extinctions associated with galactic
  dust lanes are $A_{\rm V}$$\sim$0.5-1.5 mag. Such level of
  extinction corresponds to gas column densities of $N_{\rm
    H}$$<$${\rm 10^{23}\,cm^{-2}}$ for gas-to-dust ratios typical of
  nearby AGN (\citealt{maiolino01}). For example the column density
  towards the Galactic Center associated with molecular gas is $N_{\rm
    H}$$\sim$ a few\,$\times$${\rm 10^{22}\,cm^{-2}}$
  (\citealt{sanders84}).}. As we only have three type 1 objects with
${\rm N_H}$$>$${\rm 10^{23}\,cm^{-2}}$ in \buxs the distribution of
\cv for such objects may not be representative of the overall
population of highly absorbed type 1 AGN. Thus, in what follows we
restrict our discussion to type 2 objects. Nevertheless, we note that,
based on the KS test and our simulation analysis, we cannot reject the
null hypothesis that the distributions of \cv for unabsorbed, mildly
absorbed (${\rm 4 \times 10^{21}\,cm^{-2}}$$<$${\rm N_H}$$<$${\rm
  10^{23}\,cm^{-2}}$ ) and highly absorbed (${\rm
  10^{23}\,cm^{-2}}$$<$${\rm N_H}$$<$$ 4 \times {\rm
  10^{23}\,cm^{-2}}$) type 1 AGN are drawn from the same parent
population. We find that the covering factor of a typical type 2 AGN
torus increases with ${\rm N_{H}}$ (Fig.~\ref{fig4} bottom
plots). This effect becomes more pronounced at column densities
approaching the Compton-thick regime.

Our analysis demonstrates that, not only AGN with different optical
classifications have on average tori with different covering factors,
but also that the most highly absorbed type 2 AGN have the highest
covering factors of nuclear dust. Since all sources with absorbing
column densities ${\rm N_H}$$>$4$\times$${\rm 10^{23}\,cm^{-2}}$ have
remarkably similar distributions of \cv it is highly unlikely that
statistical fluctuations associated with the small sample size affect
our results. Such a high dust covering factors seem to be a common
property of the most absorbed Compton-thin type 2 AGN (but see also
\citealt{silva04}).

Interestingly, \citet{ricci11} found that, type 2 AGN with column
densities in the range ${\rm 10^{23}\,cm^{-2}\leq
  N_H<10^{24}\,cm^{-2}}$ have a stronger X-ray reflection component
than both type 1 and type 2 AGN with ${\rm N_H<10^{23}\,cm^{-2}}$. If
the material in the torus is the main X-ray reflector, these results
are consistent with a scenario where the covering factor of the torus
is higher in the most highly absorbed Compton-thin type 2 AGN. This is
supported by our findings.

X-ray spectral variability studies have shown that a large fraction of
the X-ray absorbing cold gas must be located at the physical scales of
the broad line region, probably in dust-free clouds in the innermost
part of the torus, inside the dust sublimation radius
(e.g. \citealt{elitzur08}; \citealt{risaliti09}; \citealt{bianchi12};
\citealt{markowitz14}, and references therein). Our study supports
these results, as the relationship between \cv and ${\rm N_H}$ that we
find implies that the dust and most of the X-ray absorbing gas are at
least geometrically related and plausibly belong to the same
structure, the putative torus.

\subsection{\cv versus AGN luminosity}
\label{f2Lx}
Numerous works in the literature find that the relative fraction of
type 2 AGN decreases with increasing AGN luminosity
(e.g. \citealt{hasinger05}; \citealt{ceca08}; \citealt{treister08};
\citealt{ebrero09}; \citealt{burlon11}; \citealt{ueda14};
\citealt{assef15}; \citealt{buchner15}; \citealt{lacy15}). These
results have often been interpreted in the framework of the `receding
torus' model. According to this model the radius at which the dust
sublimates increases with AGN luminosity, resulting in an increase of
the opening angle of the torus and a decrease of its geometrical
covering factor (\citealt{lawrence91}). The end result is that the
probability of finding an AGN as optical type 2 is lower at high AGN
luminosities.

\begin{figure*}
  \centering
  \begin{tabular}{cc}
    \hspace{-0.7cm}\includegraphics[angle=90,width=0.5\textwidth]{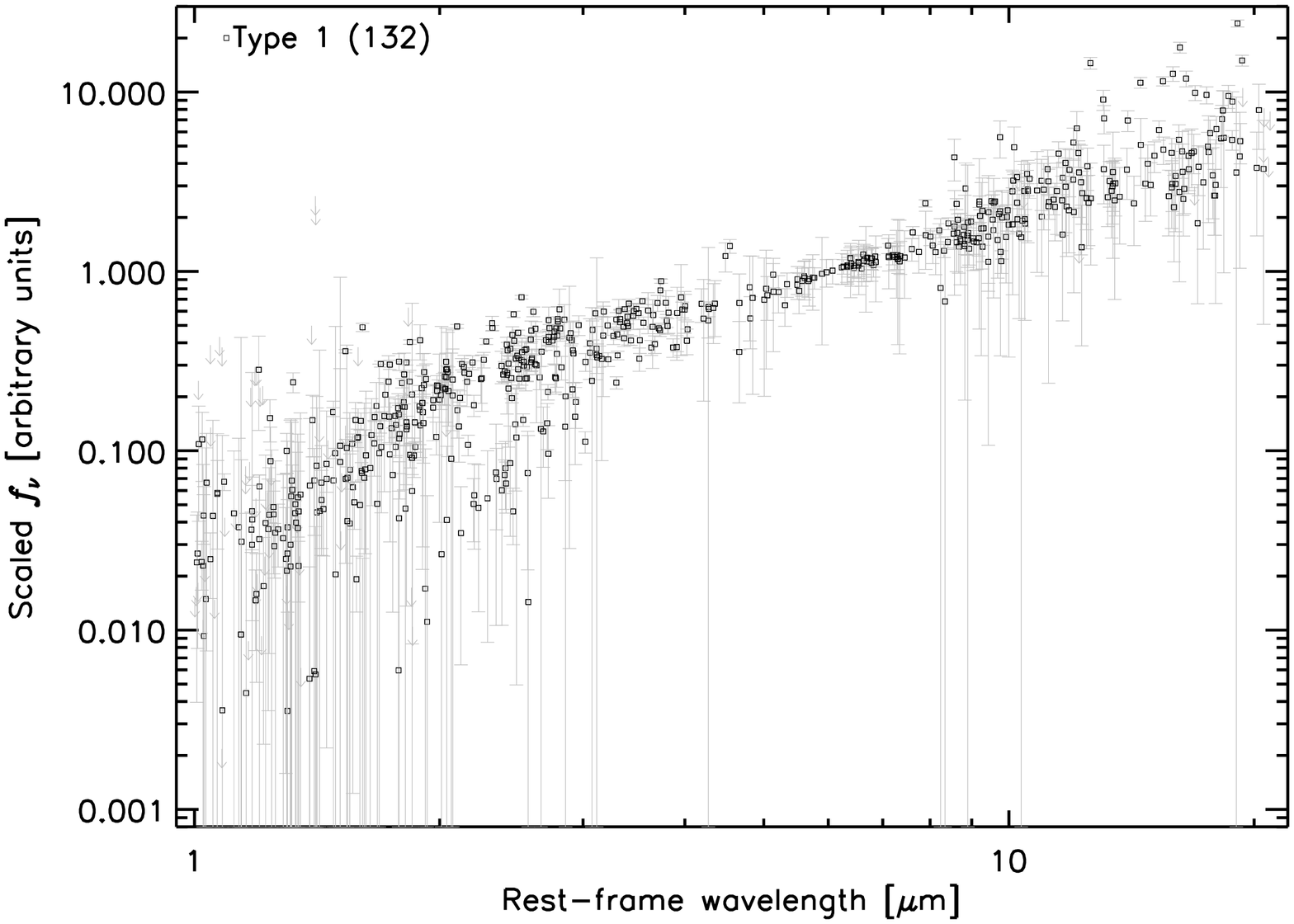}
    \hspace{-0.3cm}\includegraphics[angle=90,width=0.5\textwidth]{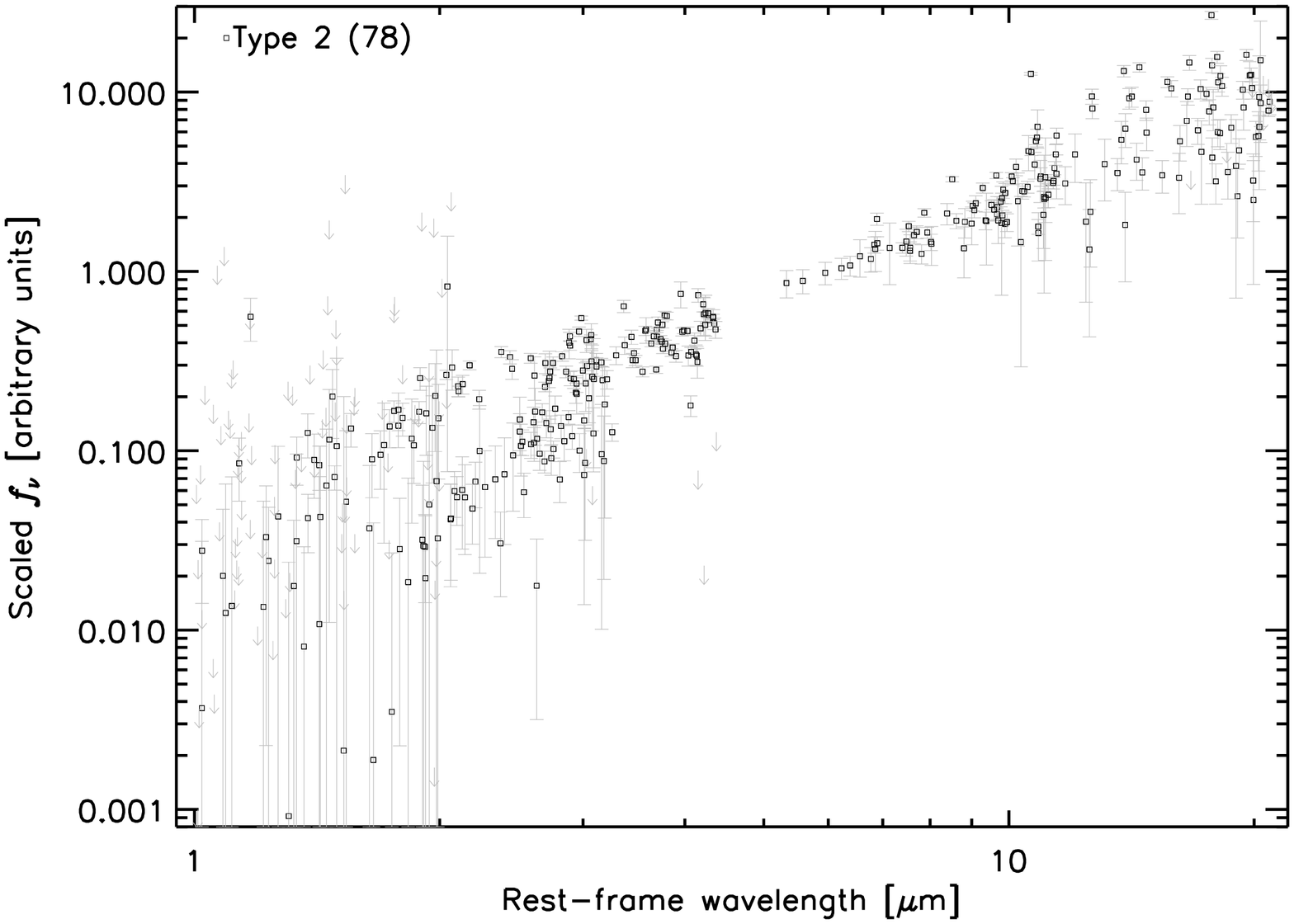}\\
    \hspace{-0.7cm}\includegraphics[angle=90,width=0.5\textwidth]{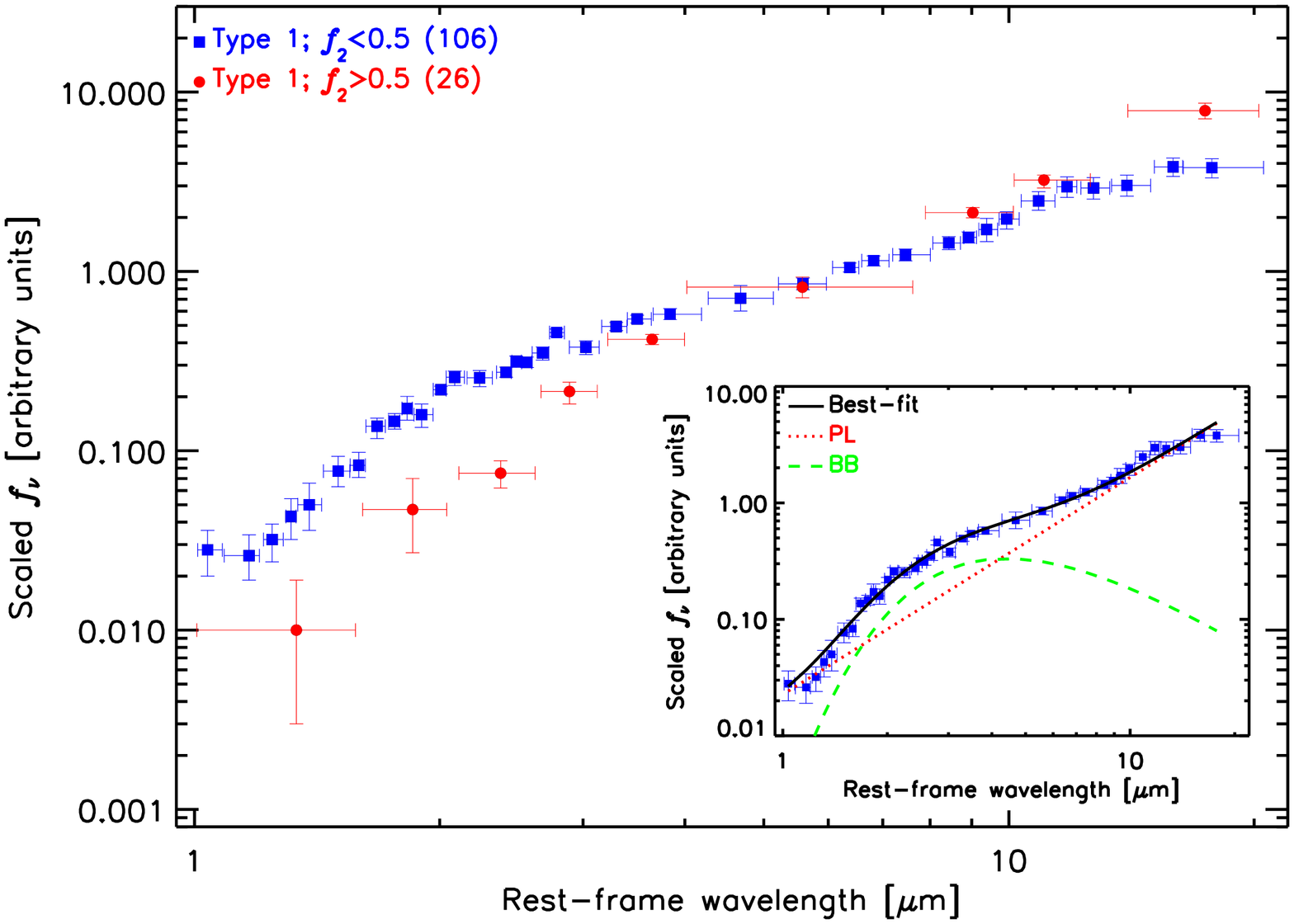}
    \hspace{-0.3cm}\includegraphics[angle=90,width=0.5\textwidth]{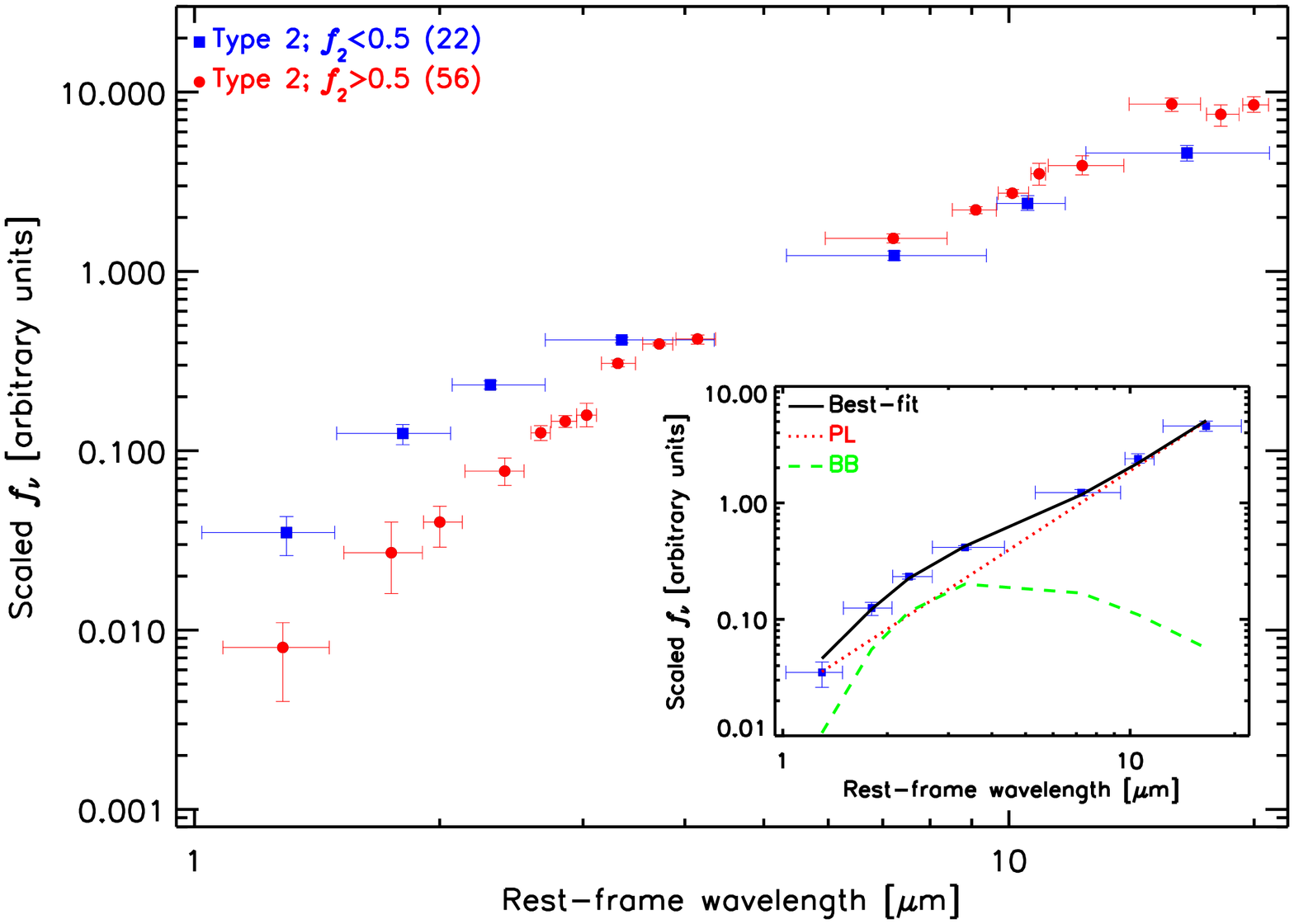}\\
  \end{tabular}
  \caption{Top plots: nuclear torus-only infrared SEDs of our type 1
    (left) and type 2 (right) AGN, respectively (small open symbols).
    Downward arrows represent upper limits. All SEDs are normalized at
    rest-frame 6\,\mic. Bottom plots: median SEDs of objects with low
    (\ncv$<$0.5; large filled squares) and high (\ncv$>$0.5; large filled
    circles) dust covering factors, respectively. The insets show the
    fits of the median SEDs of type 1 and type 2 objects with
    \ncv$<$0.5 with a two-component phenomenological model consisting
    on a power-law (PL; dotted line) and a black-body (BB; dashed
    line). Solid lines represent the best-fit model (see
    Sec.~\ref{ir_cont_f2_excess} for details). }
  \label{fig6}
\end{figure*}

To investigate whether we find any evidence supporting a scenario
where the AGN radiation field can affect the torus properties, we have
determined the distributions of \cv in two luminosity bins. The
results are illustrated in Fig.~\ref{fig5}. We clearly see that low
covering factors are preferred in type 1s at high AGN
luminosities. Although a similar trend is detected in type 2 objects,
it is much less significant. This is somewhat expected as a decrease
in the torus covering factor reduces the probability of identifying an
AGN as optical type 2. We have compared the distributions of \cv
obtained at low and high AGN luminosities for the same class of
objects using the KS test. We can reject the null hypothesis that the
two samples are drawn from the same parent population only for type 1
AGN with a significance of 99.91 per cent.

Our sample of type 1 AGN reaches $z$$\sim$1.7 while all type 2 AGN
have $z$$\lesssim$1. To avoid comparing objects at different
evolutionary stages, we have also determined the distributions of \cv
in our two luminosity bins using type 1 and type 2 objects at
$z$$<$1. Our results do not change. We note that we reach the same
conclusion adopting even lower redshift limits. This shows that,
although the rest-frame infrared spectral regions sampled with our
photometric data vary with the objects' redshift, this has no effect
on the \cv distributions presented in Fig.~\ref{fig5}. Therefore,
we can conclude that our results are robust and they are consistent
with a decrease of the covering factor of AGN tori with increasing AGN
luminosity, although for X-ray selected type 2 AGN, the effect is
modest. A detailed investigation of whether the detected changes of
\cv with luminosity are strong enough to explain the observed scarcity
of type 2 AGN at high luminosities will be presented in a forthcoming
paper.

\subsection{Nuclear infrared continuum emission}
\label{ir_cont}
\subsubsection{Dependence on \cv} \label{ir_cont_f2_dep}
It is reasonable to expect that the properties of the nuclear infrared
continuum emission of AGN, in particular the broad-band continuum
shape, might depend directly on the covering factor of the torus. To
investigate this issue we show in Fig.~\ref{fig6} (top plots) a
compilation of all nuclear infrared SEDs of our AGN\footnote{The upper
  limits indicate cases where, based on our SED decomposition
  analysis, the AGN accretion disk and host galaxy emission account
  for the full observed flux. In such cases the upper limit is the
  combined error of the fluxes associated with the accretion disk and
  host galaxy emission.}. As we are only interested in examining the
continuum shape, we have normalized all SEDs at rest-frame 6\,\mic\,
to facilitate the comparison. To do so we have used linear
interpolation in log-log space. We stress that we are only interested
in the emission associated with the dusty torus, hence as indicated
before, our nuclear infrared SEDs have been corrected for any
contamination from the extrapolated accretion disk emission and, of
course, the host galaxy.

The first result from Fig.~\ref{fig6} is that there is a large range
of torus-only infrared continuum shapes for both type 1 and type 2
AGN. Clearly, at rest-frame wavelengths shorter than $\sim$20\,\mic\,
there is no canonical infrared slope for either AGN class (for similar
results see also \citealt{alonso03}; \citealt{lira13}). More
importantly, we find that type 1 and type 2 AGN show a similar range
of infrared continuum slopes although on average type 2 AGN have
steeper SEDs.

To investigate the role of the torus covering factor we have compared
the SEDs of objects with tori with low (\ncv$<$0.5) and high
(\ncv$>$0.5) covering factors, respectively. To assign the objects to
the \ncv$<$0.5 or \ncv$>$0.5 class we have used the median values of the
posterior distributions of \cv obtained with the Bayes{\sc CLUMPY} SED
fits for each source. We then moved all individual SEDs to rest-frame
wavelengths and distributed the photometric data points into a common
wavelength grid. The bins were defined to have at least 13 points and
a minimum size of 0.01\,\mic. We have used the Astronomy Survival
Analysis package (ASURV; \citealt{isobe86}) to determine the median
flux values in each bin taking into account both detections and
upper-limits. To determine the errors of the median SEDs fully taking
into account both the dispersion and errors in the individual fluxes
we used Monte Carlo simulations. For each photometric point, if it was
an upper-limit, we kept the values unchanged while for detections we
generated random numbers using Gaussian distributions of mean and
sigma, the flux measurements and their corresponding uncertainties,
respectively. In cases where the simulated flux values were lower than
zero, we replaced them with the corresponding flux uncertainties and
treated then as upper-limits. We repeated the Monte Carlo exercise
10$^{\rm 4}$ times calculating each time median fluxes in the bins
using the ASURV package, as we did for the real data. We then
determined the uncertainties in our median SED fluxes using the 16th
and 84th percentiles (68 per cent enclosed, equivalent to 1$\sigma$)
of the distributions of simulated flux values on each bin. The results
of this analysis are illustrated in Fig.~\ref{fig6} (bottom plots).

\begin{table}[!t]
 \caption{Broad-band continuum shape of the observed and nuclear infrared
   emission of AGN.}
 \label{tab:alpha_ir}
 \centering
 \begin{tabular}{@{}ccccccccc}
   \hline
   \hline
  Class & $f_{\rm 2}$ & N & $\langle$$L_X$$\rangle$ & $\alpha_{\rm torus}$ & $\alpha_{\rm torus+disk}$ & $\alpha_{\rm obs}$\\
   (1)   &  (2)  & (3) & (4) & (5) & (6) & (7)\\
  \hline 
  Type\,1 & All    & 132 & 44.34 & 1.48$\pm$0.06 & 1.34$\pm$0.07  & 1.22$\pm$0.07 \\
  Type\,1 & $<$0.5 & 106 & 44.41 & 1.40$\pm$0.08 & 1.27$\pm$0.08  & 1.14$\pm$0.08  \\
  Type\,1 & $>$0.5 &  26 & 43.74 & 1.99$\pm$0.13 & 1.70$\pm$0.14  & 1.58$\pm$0.11  \\ \\

  Type\,2 & All    & 78  & 43.49 & 1.80$\pm$0.07 & 1.74$\pm$0.07  & 1.47$\pm$0.08 \\
  Type\,2 & $<$0.5 & 22  & 43.95 & 1.61$\pm$0.13 & 1.55$\pm$0.14  & 1.23$\pm$0.14  \\
  Type\,2 & $>$0.5 & 56  & 43.46 & 1.86$\pm$0.08 & 1.81$\pm$0.09  & 1.52$\pm$0.10  \\
  \hline
 \end{tabular}
 $Notes$: (1): optical spectroscopic classification; (2): interval of
 torus geometrical covering factors of the sample; (3): number of
 objects; (4): median X-ray luminosity in logarithmic units; (5), (6)
 and (7): power-law indices of the median infrared continuum at
 rest-frame 5-20\,\mic\, in the following cases: the SEDs
 include only the emission associated with the torus (column 5); the
 SEDs include the emission associated with both the torus and the
 accretion disk (column 6); the SEDs include the emission associated
 with both the torus and the accretion disk and have not been
 corrected for any contamination associated with the host galaxies
 (column 7).
 \smallskip
 \end{table}
 \smallskip

We clearly see that AGN with high \cv have significantly redder SEDs
on average than those with low \ncv. We find the same result for both
type 1 and type 2 objects. As indicated in Table~\ref{tab:alpha_ir},
type 1 (type 2) AGN with tori with low covering factors are on average
about 5 (3) times more luminous than those with high covering
factors. The observed differences in their nuclear infrared continuum
emission could be a manifestation of the decrease of \cv with AGN
luminosity (see Sec.~\ref{f2Lx}).

Interestingly, we find that $\sim$20 per cent of type 1 AGN have tori
with large, \ncv$>$0.5, covering factors while $\sim$28 per cent of
type 2 AGN have tori with small, \ncv$<$0.5, covering factors. As
pointed out in Sec.~\ref{agn_sample}, Seyfert 1.9 objects can only be
identified up to $z$$\lesssim$0.4 hence, some might still be present
in our sample of type 2 AGN. To evaluate whether this has any impact
on our results we have determined the fraction of type 2 AGN at
$z$$<$0.3 that have tori with \ncv$<$0.5, finding a value of 22.7 per
cent (10 out of 44 objects). We can safely conclude that $\sim$23-28
per cent of type 2 AGN have tori with small, \ncv$<$0.5, covering
factors.

To investigate whether differences exist in the continuum emission of
AGN tori as a function of \cv across the full range of wavelengths
sampled, we have determined the spectral index that best describes the
broad-band continuum emission of AGN tori at rest-frame 5-20\,\mic. We
have used a phenomenological model consisting of a simple
power-law\footnote{We characterize the rest-frame 5-20\,\mic\,
  continuum as $f_\nu$\,$\propto$\,$\nu^{-\alpha}$, where $\alpha$ is
  the power-law index, $\nu$ are frequencies and $f_\nu$ are flux
  densities, respectively.}  which provides a good description of the
data at these wavelengths. The results are shown in column 5 in
Table~\ref{tab:alpha_ir}. Although the numbers are broadly consistent,
within the uncertainties, we find that type 2 AGN have on average
steeper spectral indices than type 1 AGN. This is expected as, even
within the chosen \cv bins, the former AGN class has tori with higher
covering factors overall than the later. Clearly, at rest-frame
wavelengths longer that 5\,\mic\, differences still exist in the
continuum emission of AGN tori, even among AGN of the same optical
class.

Although, based on the shape of the torus continuum emission, type 1
and type 2 AGN are statistically different, there is no sharp division
between the nuclear infrared SEDs of the two AGN populations. This
implies that from the torus continuum emission alone, we cannot
unambiguously distinguish type 1 and type 2 AGN.

Based on our results we can conclude that the covering factor of the
torus is one of the main physical parameters controlling the shape of
the nuclear infrared continuum emission of AGN. Significant
differences exist in the properties of the torus emission, even among
AGN of the same optical class, implying that infrared flux-limited
population studies at rest-frame wavelengths shorter than
$\sim$5-6\,\mic\, are not free of biases against the AGN with tori
with the highest covering factors. We have shown in Sec.~\ref{f2_nh}
that these objects are on average the most highly absorbed AGN in
X-rays.

\subsubsection{Near-infrared hot dust emission} \label{ir_cont_f2_excess}
A broad near-infrared bump above the extrapolation of the rest-frame
$>$5\,\mic\, continuum is clearly detected in the SEDs of our type 1
and type 2 AGN with tori with low covering factors at rest-frame
wavelengths $\sim$1-4\,\mic. The physical origin of such feature is
still not clear. It could be associated with thermal radiation from
hot dust in the innermost part of the torus heated by the AGN
radiation field and with near sublimation temperatures (for
graphite-type and silicate-type grains) or, alternatively, it might be
emission from hot dust not associated with the torus (e.g in the
Narrow Line Region; \citealt{edelson86}; \citealt{barvainis87};
\citealt{minezaki04}; \citealt{kishimoto07}; \citealt{schweitzer08};
\citealt{mor09}; \citealt{riffel09}; \citealt{mor12}).

We have fitted the rest-frame 1-20\,\mic\, median SEDs of our type 1
and type 2 AGN with low torus covering factors with a two-component
phenomenological model consisting of a power-law and a black-body to
account for the near-infrared bump. We stress that this model is not
physically motivated, nor are we claiming that the near-infrared bump
originates in a separate component from the torus. Indeed, we find
acceptable fits for all torus-only SEDs at rest-frame 1-20\,\mic\,
with the N08 models and, after a careful visual check of the results
from Bayes{\sc CLUMPY}, we find no evidence for additional
components. The goal of our exercise is to compare the properties of
the nuclear hot dust emission in type 1 and type 2 AGN. The results of
the fits are illustrated in Fig.~\ref{fig6} (insets in the bottom
plots).

The values obtained for the mean spectral indices of the mid-infrared
broad-band continuum and black-body temperatures are
$\alpha$=1.87$\pm$0.07 and T=1154.2$\pm$33.2\,K for type 1 AGN and
$\alpha$=1.95$\pm$0.12 and T=1180.1$\pm$81.2\,K for type 2 AGN. The
best-fit temperatures indicate emission from very hot dust close to
sublimation temperature. To determine the strength of the
near-infrared bump we have computed its relative contribution to the
integrated flux at rest-frame 2-7\,\mic. We found a contribution of
49.5$\pm$3.4 per cent in type 1 AGN and 41.6$_{-7.8}^{+6.5}$ per cent
in type 2 AGN, respectively. Clearly, not only is the near-infrared
bump not exclusively detected in type 1 AGN, but it also appears to
have the same overall shape in type 1 and type 2 AGN with tori with
low covering factors.

\subsubsection{Contamination from the accretion disk and AGN hosts} \label{ir_cont_f2_cont}
So far, we have analysed nuclear infrared SEDs corrected for
contamination from the extrapolated accretion disk emission and the
AGN host galaxies. To compare our results with previous studies in the
literature, which normally do not apply these corrections, we have
analysed the median AGN SEDs that also include the emission from the
accretion disk, and the median AGN SEDs based on the catalogued
photometry, that include also the host galaxy emission. To do so we
have followed the same approach as in Sec.~\ref{ir_cont_f2_dep},
fitting the rest-frame 5-20\,\mic\, continuum emission with a simple
power-law. The results are summarized in columns 6 and 7 in
Table~\ref{tab:alpha_ir}. Only when we used median SEDs based on
catalogued fluxes did we obtain spectral indices consistent with the
typical values reported in the literature, especially for type 1 AGN
(e.g. \citealt{alonso06}; \citealt{buchanan06}; \citealt{hernan09};
\citealt{wu09}; \citealt{mullaney11}). This demonstrates that not only
the emission from the accretion disk but also the stellar emission
from the hosts can have a significant impact on the measured best-fit
spectral indices of the infrared emission of AGN tori.

\section{Discussion and conclusions}
Our study aims to test AGN unified models in the framework of clumpy
torus models. To do so we have determined the distribution of dust
covering factors of AGN tori using a large, uniformly selected,
complete sample of 227 AGN. The AGN belong to the Bright Ultra-hard
\xmm Survey and have $z$ in the range 0.05-1.7, and 2-10 keV intrinsic
(absorption-corrected) luminosities between 10${\rm ^{42}}$ and
10${\rm ^{46}}$ $\lum$.

Employing data from UKIDSS, 2MASS and WISE and a thorough SED
decomposition analysis into AGN and galaxy emission, in a previous
paper we determined the rest-frame 1-20\,\mic\, continuum emission
associated with the torus for our sample objects. Here we modelled our
nuclear infrared SEDs with the clumpy torus models of
\citet{nenkova08} using the code Bayes{\sc CLUMPY}. This program has
been especially developed to analyze the emission of AGN tori with the
\citet{nenkova08} models using a Bayesian inference approach.

The main results of our analysis can be summarized as follows:
\begin{enumerate}
\item Type 1, type 2 and probably also intermediate-type AGN, are
  on average intrinsically different. Type 2 AGN have tori with
  higher geometrical covering factors \cv on average than type 1
  AGN. Nevertheless, the distributions of \cv for both type 1 and type
  2 AGN are broad and there is a large overlap between the two
  populations. Although rare among all AGN, we find type 1 objects
  with large torus covering factors (26 out of 132) and type 2 objects
  with small torus covering factors (22 out of 78).

\item Interestingly, type 1 and type 2 AGN with similar levels of
  X-ray absorption have significantly different distributions of
  torus geometrical covering factors. This result also holds for X-ray
  unabsorbed type 1 and type 2 objects.

  \item AGN classified as Sy1.8-1.9 have a rather flat distribution of
    \cv that is significantly different from those of type 1 and type
    2 AGN. Taking into account that most Sy1.8-1.9s are absorbed in
    X-rays (12 out of 17 objects) it is unlikely that all Sy1.8-1.9
    are simply ordinary type 1 AGN caught in a low flux state during
    the UV/optical spectroscopic observations. Our results favour a
    scenario where most Sy1.8-1.9s have broad-line regions reddened by
    optically-thin dust located either in the torus or on physical
    scales of the narrow-line region or the host galaxies.

\item \cv increases with the X-ray column density, at least at ${\rm
  N_H}$$>$${\rm 10^{23}\,cm^{-2}}$, which implies that dust extinction
  and X-ray absorption are geometrically related and plausibly belong
  to the same structure, the putative dusty torus.

\item Low \cv values are preferred at high AGN luminosities, as
  postulated by simple receding torus models, although for X-ray
  selected type 2 AGN the effect is certainly small.

\item Based on our results, \cv is one of the main physical parameters
  controlling the shape of the nuclear infrared emission of
  AGN. Although, from the shape of the torus continuum emission, type
  1 and type 2 AGN are statistically different, there is no sharp
  division between the nuclear infrared SEDs of the two AGN
  populations. This implies that from the torus continuum emission
  alone, we cannot unambiguously distinguish type 1 and type 2 AGN.

\item A broad near-infrared bump at rest-frame $\sim$1-4\,\mic\, above
  the extrapolation of the rest-frame $>$5\,\mic\, infrared continuum
  is clearly detected in the SEDs of our type 1 and type 2 AGN having
  tori with low covering factors. We find that such spectral feature,
  which is often assumed to be due to hot dust in the inner-most part
  of the torus, is not exclusively detected in type 1 AGN and it has
  the same average properties in type 1 and type 2 AGN.
\end{enumerate}

Based on the results presented here, all AGN are not intrinsically the
same. This result applies not only to AGN with different optical
classifications, but also to objects of the same optical class, in
agreement with predictions from clumpy torus models. The AGN radiation
field can modify the covering factor of the nuclear dust obscuring the
central engine, although, at least in X-ray selected type 2 objects,
the effect is rather small. Furthermore, the covering factor of the
torus increases with the X-ray column density, already for X-ray
obscuration in the Compton-thin regime.

The reported significant differences in the torus emission, even among
AGN of the same optical class, imply that infrared flux-limited
population studies at rest-frame wavelengths shorter than
$\sim$5-6\,\mic, are not free of biases against the most highly
absorbed AGN, which we have shown are the objects with tori with the
highest covering factors.

We can conclude that, the viewing angle, AGN luminosity and also \cv
determine the optical appearance of an AGN. Furthermore, \cv controls
the overall shape of the nuclear infrared continuum emission at
rest-frame $\sim$1-20\,\mic. Thus, the geometrical covering factor of
the dusty torus must be incorporated as a key ingredient of
unification schemes.

\bigskip 
We thank Nicol\'as Cardiel, R. Della Ceca and P. Severgnini for useful
comments and discussions. This work is based on observations obtained
with XMM-Newton, an ESA science mission with instruments and
contributions directly funded by ESA Member States and NASA. Based on
data from the Wide-field Infrared Survey Explorer, which is a joint
project of the University of California, Los Angeles, and the Jet
Propulsion Laboratory/California Institute of Technology, funded by
the National Aeronautics and Space Administration. Funding for the
SDSS and SDSS-II has been provided by the Alfred P. Sloan Foundation,
the Participating Institutions, the National Science Foundation, the
U.S. Department of Energy, the National Aeronautics and Space
Administration, the Japanese Monbukagakusho, the Max Planck Society
and the Higher Education Funding Council for England. The SDSS Web
Site is http://www.sdss.org/. Based on observations collected at the
European Organization for Astronomical Research in the Southern
hemisphere, Chile, programme IDs 084.A-0828, 086.A-0612, 087.A-0447
and 088.A-0628. Based on observations made with the William Herschel
Telescope and its service programme - operated by the Isaac Newton
Group, the Telescopio Nazionale Galileo - operated by the Centro
Galileo Galilei and the Gran Telescopio de Canarias installed in the
Spanish Observatorio del Roque de los Muchachos of the Instituto de
Astrof\'isica de Canarias, in the island of La Palma. SM, FJC, XB,
A.H.-C. and A.A.-H. acknowledge financial support by the Spanish
Ministry of Economy and Competitiveness through grant AYA2012-31447,
which is partly funded by the FEDER programme. SM, FJC and
A.A.-H. acknowledge financial support from the ARCHES project (7th
Framework of the European Union, No. 313146). C.R.A. acknowledges
financial support from the Marie Curie Intra European Fellowship
within the 7th European Community Framework Programme (PIEF-GA-2012-
327934). AAR acknowledges financial support through the Ram\'on y
Cajal fellowship and projects AYA2014-60476-P and Consolider-Ingenio
2010 CSD2009-00038 from the Spanish Ministry of Economy and
Competitiveness. We thank the referee for the revision of the paper.





\end{document}